\documentclass{WileyMSP-template}
\usepackage{amsthm}
\usepackage{amsmath}
\usepackage{amssymb}
\usepackage{amsfonts}
\usepackage{epsfig}
\usepackage{bm}
\usepackage{times}
\usepackage[colorlinks,linkcolor=blue,anchorcolor=blue,citecolor=blue,urlcolor=blue]{hyperref}
\usepackage{braket}
\usepackage{graphicx}
\usepackage{setspace}
\usepackage{subfigure}
\usepackage{enumerate}
\usepackage{changes}
\usepackage{color}
\usepackage{multirow}

\begin{document}

\pagestyle{fancy}
\rhead{\includegraphics[width=2.5cm]{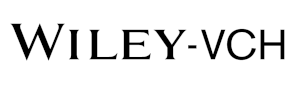}}

\title{Purity-Assisted Zero-Noise Extrapolation for Quantum Error Mitigation}

\maketitle


\author{Tian-Ren Jin}
\author{Yun-Hao Shi}
\author{Zheng-An Wang}
\author{Tian-Ming Li}
\author{Kai Xu*}
\author{Heng Fan*}



\begin{affiliations}
T.-R. Jin, Dr. Y.-H. Shi, T.-M. Li, Prof. K. Xu, Prof. H. Fan\\
Institute of Physics, Chinese Academy of Sciences, Beijing 100190, China\\
School of Physical Sciences, University of Chinese Academy of Sciences, Beijing 100049, China\\
Email Address: kaixu@iphy.ac.cn, hfan@iphy.ac.cn

Dr. Z.-A. Wang, Prof. K. Xu, Prof. H. Fan\\
Beijing Academy of Quantum Information Sciences, Beijing 100193, China\\
Hefei National Laboratory, Hefei 230088, China

Prof. K. Xu, Prof. H. Fan\\
Songshan Lake Materials Laboratory, Dongguan 523808, China\\
CAS Center for Excellence in Topological Quantum Computation, UCAS, Beijing 100190, China

\end{affiliations}


\keywords{quantum error mitigation, zero-noise extrapolation, purity, Pauli twirling, quantum computation platform, superconducting qubits}

\begin{abstract}

    Quantum error mitigation aims to reduce errors in quantum systems and improve accuracy. Zero-noise extrapolation (ZNE) is a commonly used method, where noise is amplified, and the target expectation is extrapolated to a noise-free point. However, ZNE relies on assumptions about error rates based on the error model. In this study, a purity-assisted zero-noise extrapolation (pZNE) method is utilized to address limitations in error rate assumptions and enhance the extrapolation process. The pZNE is based on the Pauli diagonal error model implemented using the Pauli twirling technique. Although this method does not significantly reduce the bias of routine ZNE, it extends its effectiveness to a wider range of error rates where routine ZNE may face limitations. In addition, the practicality of the pZNE method is verified through numerical simulations and experiments on the online quantum computation platform, \emph{Quafu}. Comparisons with routine ZNE and virtual distillation methods show that biases in extrapolation methods increase with error rates and may become divergent at high error rates. The bias of pZNE is slightly lower than routine ZNE, while its error rate threshold surpasses that of routine ZNE. Furthermore, for full density matrix information, the pZNE method is more efficient than the routine ZNE.	

\end{abstract}


\section{Introduction}
Quantum computation is expected to surpass classical computation for specific problems; nevertheless, the vulnerability of qubits to environmental noise poses a significant challenge to realizing a practical quantum computer. Consequently, the development of fault-tolerant quantum computation methodologies is crucial for enabling the widespread application of quantum computation.
Quantum error correction code (QECC) provides a systematic solution to fault-tolerant quantum computation \cite{RevModPhys.87.307}, while its application requires many physical qubits and the error rate should be lower than a threshold~\cite{knill1996threshold,10.1145/258533.258579,Kitaev_1997}.
With remarkable progress~\cite{PhysRevLett.129.030501,google2023suppressing}, however, it is challenging for the state-of-the-art technique to meet the requirements for the full implementation of QECCs. Therefore, fault-tolerant quantum computation based on high-precision logic qubits is still a far-reaching task.

In the noisy intermediate-scale quantum era~\cite{Preskill2018quantumcomputingin}, quantum error mitigation (QEM) provides an alternative approach to dealing with noise. QECCs detect and correct errors that occur in the quantum process to ensure that there are no errors in the coded logic qubits. On the contrary, QEMs allow errors to occur, but use some post-processing techniques to reduce the bias between the output information of noisy quantum circuit and the ideal quantum circuit~\cite{doi:10.7566/JPSJ.90.032001,RevModPhys.95.045005}.

In recent years, several QEM schemes have been proposed, including zero-noise extrapolation (ZNE)~\cite{PhysRevX.7.021050,cai2021multi,PhysRevLett.127.270502,PhysRevLett.119.180509,PhysRevX.8.031027}, probabilistic error cancellation (PEC)~\cite{PhysRevLett.119.180509,PhysRevX.8.031027,kandala2019error,PhysRevA.104.052607}, symmetry verification methods~\cite{PhysRevA.98.062339,PhysRevLett.122.180501,cai2021quantum}, purification methods~\cite{PhysRevX.11.031057,PhysRevX.11.041036,PhysRevA.105.022427,cai2021resourceefficient}, subspace expansion~\cite{PhysRevA.95.042308,mcclean2020decoding,PRXQuantum.2.040326,PhysRevLett.129.020502}, and learning-based error mitigation~\cite{PRXQuantum.2.040330,czarnik2022improving}. These methods have been applied on different quantum computation platforms and interesting physics~\cite{kandala2019error,PhysRevLett.120.210501,kim2023scalable,PhysRevLett.128.150504,song2019quantum,van2023probabilistic,zhang2020error,PhysRevA.100.010302,PhysRevA.107.052617,doi:10.1073/pnas.2006337117,PhysRevX.8.011021,google2020hartree,arute2020observation,stanisic2022observing,PhysRevApplied.15.034026,kim2023scalable,10313813}. In addition, there have been some efforts to integrate various QEM schemes in a generalized framework~\cite{cai2021practical,cai2021quantum,takagi2022fundamental,kim2023evidence,quek2023exponentially,PhysRevLett.131.210601}. Among these various error mitigation schemes, the ZNE and PEC are used to deal with quantum circuits, which based on specific error models of noise. The PEC method mitigates the error by post-cancelling the noise, which is previously estimated from experiments in the selected error basis. Although the Pauli basis is sufficient with Pauli twirling technique~\cite{PhysRevA.94.052325,cai2019constructing,PhysRevX.11.041039,van2023probabilistic}, in practice, the selected basis should be truncated for large systems, such as the $2$-weighted Pauli basis~\cite{van2023probabilistic}. Thus, the effectiveness of PEC depends on the error model described by the error basis. Recently, an attempt was made to integrate ZNE into the PEC framework~\cite{kim2023evidence}.

However, the ZNE method is a subtle approach based on the analyticity of the observable expectation related to the error rate, which is a fundamental physical condition. The relation to the error model emerges when the error rate is considered. To perform the ZNE, the error must be amplified, and the ideal value is obtained by extrapolating the expectation backward to the point of zero noise. In the early application of ZNE, the error is amplified by stretching the duration of pulse for each gate~\cite{kandala2019error}. However, this amplification is not suitable for digital quantum gates. Another implementation is to deliberately add redundant gates in the circuit~\cite{PhysRevLett.120.210501}, which is improved with a treatment called unitary folding~\cite{9259940}. In the unitary folding method, the redundant gates are implemented by the target circuit of interest. Specifically, the circuit evolves forward and backward successively, and the error rate is characterized by the number of folds. Although the amplifications of the error are developed more feasible, the estimation of the error rate is a priori, which has limitations on the noise type. Moreover, for the complex circuit in the experiment, the effect of errors cannot be sufficiently described by a single parameter of the error rate alone.

In the ZNE method, dependence of the error model in noisy amplification is contrary to the universal idea of the extrapolation, which limits the effectiveness of this method. In this paper, we try to release the limitation of noise in the ZNE method with the assistance of the purity of the noisy output state. Here, we introduce the purity-assisted zero-noise extrapolation (pZNE) method. Specifically, we propose a modified purification method based on the Pauli diagonal error model. By estimating the bias of the modified purification methods, it is shown that the bias is controlled by error rate, which can be represented by purity of noisy state. The pZNE method mitigates the noisy expectation by extrapolating the mitigated expectations of the modified purification method with different error rates to a noise-free pure state.
Equivalently, it can be applied by extrapolating the noisy expectations with purity.

This paper is organized as follows. In Sec.~\ref{sec: ZNE}, we review the routine ZNE method and analyze how this method is based on the error model of noise. Then, we show that the purity of the output state can assist the extrapolation in zero-noise error mitigation in Sec.~\ref{sec: purity}. We describe the detailed procedures of pZNE in Sec.~\ref{sec: detail}, including the fitting model, purity estimation methods, and the measurement overhead. We also verify this method by numerical simulations and experiments on \emph{Quafu} cloud-based quantum computation platform in Sec.~\ref{sec: simu}. The conclusion and discussion are given in Sec.~\ref{sec: conclusion}.

\section{Zero-noise Extrapolation with Unitary Folding} \label{sec: ZNE}

The ZNE method consists of two steps:
\begin{enumerate}
    \item \textbf{Noise amplification}: Collecting the raw data of different error rates via measuring on modulated circuits.
    \item \textbf{Extrapolation}: Post-processing the experiment data to obtain the mitigated expectation value based on some specific fitting models.
\end{enumerate}
In this section, we will review the ZNE method, the fitting models, and the noise amplification method of unitary folding.
In addition, we discuss how the assumption that the error rate is proportional to the number of folds is based on the error model of noise.

Consider the ideal unitary circuit $\mathcal{U}$ of interest. However, in practice, we can only perform a noisy circuit $\mathcal{U}_{\lambda}$ instead.
Then the experimental expectation value $\braket{\hat{O}}_{\lambda} = \mathrm{Tr}[\hat{O} \rho_{\lambda}]$ of observable $\hat{O}$ of interest is different from the ideal one $\braket{\hat{O}}_{0} = \mathrm{Tr}[\hat{O} \rho_{0}]$, where $\rho_{\lambda} = \mathcal{U}_{\lambda}(\rho)$ is the noisy output state, and $\rho_{0} = \mathcal{U}(\rho)$ is the ideal output state.
Now, we are going to infer the ideal value $\braket{\hat{O}}_{0}$ from some noisy expectation values $\braket{\hat{O}}_{\lambda}$.

This method is inspired by a simple idea that the expectation value $\braket{\hat{O}}_{\lambda}$ is an analytic function of a parameter $\lambda \geq 0$, which characterizes the noise level of noisy circuit $\mathcal{U}_{\lambda}$, and assume $\braket{\hat{O}}_{\lambda} = \braket{\hat{O}}_{0}$ when $\lambda = 0$.
According to analyticity, the expectation value can be expanded in a power series of $\lambda$ as
\begin{equation}
    \braket{\hat{O}}_{\lambda} = O(\lambda) = \sum_{n = 0}^{\infty} O_n \lambda^n,
\end{equation}
and by definition of $\lambda$, we have $O_0 = O(0) = \braket{\hat{O}}_{0}$.
Thus, by interpolation the function $O(\lambda)$ with expectations $\braket{\hat{O}}_{\lambda_i}$ of different error rate $\lambda_i$, one can infer the noise-free expectation value $\braket{\hat{O}}_{0}$.

For the number of experimental data is finite, the expansion of $O(\lambda)$ should be truncated at some order $M$ of $\lambda$, which requires that the error rate $\lambda$ should small enough, and this is the polynomial model of function $O(\lambda)$~\cite{PhysRevX.7.021050,PhysRevLett.119.180509}.
In addition, there are also (multi-)exponential model~\cite{cai2021multi}
\begin{equation}
    \braket{\hat{O}}_{\lambda} = O(\lambda) = \sum_{a = 0}^{M} B_a e^{- \gamma_a \lambda},
\end{equation}
and poly-exponential model~\cite{9259940}
\begin{equation}
    \braket{\hat{O}}_{\lambda} = O(\lambda) = e^{f(\lambda)} + O_{\infty},
\end{equation}
where $f(\lambda)$ is a polynomial of $\lambda$.
These fitting models are put forward under the consideration that the physical observable should be bounded.

\begin{figure}[t]
    \centering
    \includegraphics[width=0.45\textwidth]{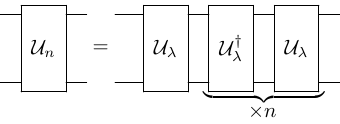}
\caption{The circuit of unitary folding.}
    \label{fig: unitary}
\end{figure}

Then, the noisy data of different error rates $\lambda_i$ is obtained by error amplification.
The unitary folding performs the circuit sequence in the experiment
\begin{equation}
    \mathcal{U}_{n} = \mathcal{U}_{\lambda} \circ (\mathcal{U}_{\lambda}^{\dagger} \circ \mathcal{U}_{\lambda})^{n-1} = (\mathcal{U}_{\lambda} \circ \mathcal{U}_{\lambda}^{\dagger})^{n-1} \circ \mathcal{U}_{\lambda},
\end{equation}
where $\circ$ denotes the composition of quantum operations, the corresponding error rate is expected as $\lambda_n = (2 n + 1) \lambda_0$, and the resolution is $2 \lambda_0$.
The circuit is shown in Figure~\ref{fig: unitary}.
To have a more fine-grained resolution, it can also perform the layer (gate) folding~\cite{9259940}, which assumes the gate can be decomposed into layers $\mathcal{U}_{\lambda} = \prod_{i=1}^{d} \circ \mathcal{L}_i$, and performs the folding on these layers
\begin{equation}
    \mathcal{L}_i \mapsto (\mathcal{L}_i \circ \mathcal{L}_i^{\dagger})^n \circ \mathcal{L}_i
\end{equation}
in the circuit.
By doing so, the resolution can be promoted to $2 \lambda/d$.

To consider the procedure of unitary folding in more detail, we assume that the circuit is a gate only, and set the noise in forward and backward evolutions as
\begin{align}
    \mathcal{U}_{\lambda} & = \mathcal{E}_f \circ \mathcal{U}, \\
    \mathcal{U}_{\lambda}^{\dagger} & = \mathcal{U}^{\dagger} \circ \mathcal{E}_b,
\end{align}
which can be interpreted as the definition of forward error $\mathcal{E}_f$ and backward error $\mathcal{E}_b$. The error rate for both errors is denoted as $\lambda_f$ and $\lambda_b$ respectively.
Then, we formally have
\begin{equation}
    \mathcal{U}_{\lambda} \circ \mathcal{U}_{\lambda}^{\dagger} = \mathcal{E}_f \circ \mathcal{E}_b \equiv \mathcal{E},
\end{equation}
thus the noisy gate is
\begin{equation}
    \mathcal{U}_{n} = \mathcal{E}_{n} \circ \mathcal{U},
\end{equation}
where $\mathcal{E}_{n} = \mathcal{E}^{n-1} \circ \mathcal{E}_f$ is the error channel with the error rate $\lambda_n = (n-1) \tilde{\lambda} + \lambda_f$, where $\tilde{\lambda}$ is the error rate of error channel $\mathcal{E}$.

To obtain $\lambda_n = (2 n - 1) \lambda_f$, we can assume $\tilde{\lambda} = 2 \lambda_f$, which is true when the error channel satisfies $\mathcal{E}_f = \mathcal{E}_b$.
However, this assumption is not always realized~\cite{henao2023adaptive} (or see Appendix~\ref{app: errors}).
If we assume $\tilde{\lambda} \approx 2 \lambda_f$, it will introduce an additional error in the inference of the ideal value $O(0)$ before fitting the experimental data.
As a result, the ZNE method is constrained by the error model of noise.
The direct way to solve this problem is to determine the error rate $\lambda$ from experiments.

\section{Extrapolation with Purity} \label{sec: purity}

Fidelity is always used to witness the difference between the state with an ideal state, however, the ideal state $\rho_0$ cannot be prepared precisely due to the inevitable noise in the experiment.
Therefore, measuring the fidelity of the noisy state and the ideal state requires the classical simulation of the quantum circuit, which is generally impractical for large systems.
In this section, we will describe the idea of using the purity of the output state of the noisy circuit
\begin{equation}
    p(\rho_{\lambda}) = \mathrm{Tr} |\rho_{\lambda}|^2 = \mathrm{Tr} (\rho_{\lambda}^{\dagger} \rho_{\lambda})
\end{equation}
to reflect the influence of noise in the circuit.
The ideal unitary circuit will not influence the purity of the state.
We assume that the error is not unitary, otherwise, it can be canceled by the folding of the well-calibrated circuit.
Moreover, we assume the error is in the form of Pauli diagonal 
\begin{equation}
    \mathcal{E}_{r} = \sum_{i} q_{ri} \mathcal{P}_i ,
\end{equation}  
where $\mathcal{P}_i(\cdot) = \hat{P}_i \cdot \hat{P}_i$ is Pauli operations, and $r = f, b$ distinguish the forward and backward errors.
This assumption can be realized by the Pauli twirling technique.

Let the ideal state be decomposed into Pauli basis $\rho_0 = \mathcal{U}(\rho) = \frac{1}{D}\sum_i \rho_i \hat{P}_i$.
The Pauli diagonal forward error channel acts as
\begin{equation} 
    \mathcal{E}_{r}(\rho_0) = \frac{1}{D} \sum_{i,j} q_{rj} \rho_i \mathcal{P}_j(\hat{P}_i) = \frac{1}{D} \sum_{i} \chi_{ri} \rho_i \hat{P}_i,
\end{equation} 
where $\chi_{ri} = \sum_j \epsilon_{ij} q_{rj}$ is the eigenvalue of the error, with $\mathcal{P}_j(\hat{P}_i) = \epsilon_{ij} \hat{P}_i$.
	We assume that the operators whose expectation values of interest are Pauli operators.
	Then, the expectation of $\hat{P}_i$ of the folded noisy state $\rho_{n} = \mathcal{U}_{n}(\rho) = \mathcal{E}_{n}(\rho_0)$ are
\begin{align} \label{eq: expectation}
    \braket{\hat{P}_i}_n = \mathrm{Tr}[\hat{P}_i \mathcal{E}_{n}(\rho_0)] = \chi_{fi}^n \chi_{bi}^{n-1} \rho_i 
\end{align}
By fitting these expectations with the exponential model, only the parameters $\chi_{fi} \chi_{bi}$ and $\chi_{fi} \rho_i$ are obtained.
To find the ideal expectation $\braket{\hat{P}_i}_0 = \rho_0$, we should estimate $\chi_{fi}$. 
In general, $\chi_{fi} \neq \chi_{bi}$, so the ideal expectation cannot be extracted.

We use purity to do this estimation.
The purity of the noisy state $\rho_{n}$ is
\begin{equation} \label{eq: purity}
    p_n = \frac{1}{D} \sum_i \rho_i^2 \chi_{ni}^{2} 
    = \left(p_0 - p_{\infty}\right) \overline{\chi_n^2} +  p_{\infty},
\end{equation}
where $\overline{\chi_n^2} = \frac{\sum_{\chi_{ni} < 1} \rho_i^2 \chi_{ni}^{2}}{\sum_{\chi_{ni} < 1} \rho_i^2}$ is the average of nontrivial eigenvalues $\chi_{ni} = \chi_{fi}^n \chi_{bi}^{n-1}$ of folding circuits,  $p_0 = \frac{1}{D} \sum_i \rho_i^2$ is the purity of ideal state, and $p_{\infty}$ is the purity of the stable state of the forward error channel.
For convenience, we can assume the ideal state is pure, $p_0 = 1$, and the stable state is a maximally mixed state, $p_{\infty} = 1/D$.
If the distribution of $\chi_{ni}$ concentrates, so that the variance $\Delta \overline{\chi}_n^2 = \overline{\chi_n^2} - \overline{\chi}_n^2$ is small, then we can approximate 
\begin{equation}
    \chi_{ni} \approx \overline{\chi_n^2}^{1/2} = \sqrt{\frac{p_n - p_{\infty}}{p_0 - p_{\infty}}},
\end{equation}  
and the ideal expectation is estimated as 
\begin{equation} \label{eq: estimation}
    \braket{\hat{P}_i}_0 \approx \braket{\hat{P}_i}_{n\mathrm{est}} =\braket{\hat{P}_i}_n \sqrt{\frac{p_0 - p_{\infty}}{p_n - p_{\infty}}}.
\end{equation}

The effectiveness of this approximation that $\chi_{ni} \approx \overline{\chi_n^2}^{1/2}$ requires $\Delta \overline{\chi}_n^2$ to be small, which is based mainly on the distribution of eigenvalue of error channel and slightly on the state $\rho_0$ in large systems.
The relative bias of this method is 
\begin{equation}
    \left\vert\frac{\braket{\hat{P}_i}_{n\mathrm{est}}}{\braket{\hat{P}_i}_0}- 1\right\vert = \left\vert\frac{\chi_{ni}}{\overline{\chi_n^2}^{1/2}}- 1\right\vert.
\end{equation} 
Given a tolerant error $\epsilon$, the probability of failing is 
\begin{equation} \label{eq: effectiveness}
    P\left(\overline{\chi_n^2}^{-1/2} \left\vert\chi_{ni} - \overline{\chi_n^2}^{1/2}\right\vert \geq \epsilon\right) \leq 2 e^{-\frac{\epsilon^2}{2 \sigma^2}} \cosh \frac{\epsilon}{2},
\end{equation}
where $\sigma^2 = \frac{\Delta\bar{\chi}_n^2}{\bar{\chi}_n^2}$, by Chebyshev's inequality (for detail, see Appendix~\ref{app: effectiveness}).

Denote the failing probability as $\delta$, and select the tolerant error $\epsilon$ so small, we expand the left-hand of inequality to the leading term of $\epsilon$ 
\begin{equation}
    \epsilon \leq 2 \sigma \sqrt{\frac{2 - \delta}{4  - \sigma^2}} \sim \sqrt{2} \sigma,
\end{equation} 
which means that the pZNE method works if the eigenvalues of forward error satisfy $\sigma \leq 2$, and the tolerant error $\epsilon$ is controlled by $\sigma$.
The parameter $\sigma$ relies on the form of the forward error, for example, for the global depolarizing error, all the nontrivial eigenvalues $\chi_{ni}, (i \neq 0)$ are the same, so $\sigma = 0$ no matter how large the error.
Moreover, without specifying the form of errors, this parameter is controlled by the error probability $q_{\lambda} = \sum_{i \neq 0} q_{ni}$, where $q_{ni} $ is the coefficients of Pauli diagonal error $\mathcal{E}_n$.
Obviously $1 - 2 q_{\lambda} \leq \bar{\chi}_n \leq 1$, we have
\begin{equation}
    \sigma \leq \sqrt{\frac{1 - \bar{\chi}_n}{\bar{\chi}_n}} \leq  \sqrt{\frac{2 q_{\lambda}}{1 - 2 q_{\lambda}}},
\end{equation}
thus when the error is so small that $q_{\lambda} < 0.4$, this method always has the possibility of success, and the relative bias is roughly controlled as 
\begin{equation}
    \epsilon \leq 2 \sqrt{\frac{(2 - \delta)q_{\lambda}}{2 - 5 q_{\lambda}}}.
\end{equation}
With more detail estimation (see Appendix~\ref{app: effectiveness}), $\epsilon \sim \sqrt{2} \lambda + O(\lambda^2)$.

In all, the estimator in Equation~(\ref{eq: estimation}) has bias controlled by the error probability $q_{\lambda}$ in general, so the number of folding.
Therefore, we can extrapolate the estimators of different folding circuits to reduce bias. In this step there are two strategies.
One is to extrapolate the estimators $\braket{\hat{P}_i}_{n\mathrm{est}}$ to $n = \frac{1}{2}$, which is similar to the extrapolation of routine ZNE.
Another is to extrapolate the estimators $\braket{\hat{P}_i}_{n\mathrm{est}}$ along the effective error rate $s_n = - \ln \frac{p_n - p_{\infty}}{p_0 - p_{\infty}}$ to $s_{n_0} = 0$, and in this effective error rate, 
\begin{equation}
    \braket{\hat{P}_i}_n = \braket{\hat{P}_i}_{n\mathrm{est}} e^{-\frac{1}{2} s_n} = f(s_n),
\end{equation}
which means that we can also extrapolate the noisy expectation $\braket{\hat{P}_i}_n$ with the effective error rate, or purity.

\section{Procedure of Methods} \label{sec: detail}

In Sec.~\ref{sec: purity}, we discuss the procedure of the pZNE method.
Assume our quantum computation task consists of $(\rho, \mathcal{U}, \hat{O})$, where $\rho$ is the initial state, $\mathcal{U}$ is the ideal unitary evolution, and $\hat{O}$ is the operator of interest.
The outcome we want from this computation task is the expectation $\braket{\hat{O}}_0 = \mathrm{Tr}[\hat{O} \mathcal{U}(\rho)]$.
The procedure of the pZNE method to mitigate the error in this task is following.
\begin{enumerate}
    \item \textbf{Task decomposing}: Decompose the operator $\hat{O}$ of interest into Pauli basis $\hat{O} = \sum_{i\in I} O_i \hat{P}_i$, where $I$ is the set of indices with nonzero coefficients $O_i$.
    \item \textbf{Pauli twirling}: Construct the Pauli-twirling the ideal circuit $\mathcal{U}_{\mathrm{twirl}}$, which corresponds to the noisy circuit $\mathcal{U}_{\lambda}$ in experiment.
    \item \textbf{Task performing}: Perform the circuit $\mathcal{U}_{n}$ to estimate the expectation $\braket{\hat{P}_i}_n = \mathrm{Tr}[\hat{P}_i \mathcal{U}_{n}(\rho)]$ and the purity $p_n = \mathrm{Tr}[\mathcal{U}_{n}(\rho)^2]$. 
    \item \textbf{Data processing}: Estimate $\braket{\hat{P}_i}_{n\mathrm{est}}$ with $\braket{\hat{P}_i}_n$ and $p_n$. 
    Extrapolate $\braket{\hat{P}_i}_{n\mathrm{est}}$ to noise-free point, $n = \frac{1}{2}$ or $s_{n_0} = 0$. 
    Calculate the estimation of the ideal expectation $\braket{\hat{O}}_0$ with coefficients $O_i$.
\end{enumerate} 

In this section, we will discuss the steps of task performing and data processing (for the technique of Pauli twirling, see Appendix~\ref{app: pauli_twirling}).

\subsection{Task Performing}

The unitary folding technique to construct the circuit sequence ${\mathcal{U}}_n$ is discussed in the previous section. Unlike the routine ZNE method, we use purity to assist the estimation of ideal expectations.
However, it is not an easy task to measure purity, since purity is not an expectation of linear operator on Hilbert space $\mathcal{H}$, but $\mathcal{H}^{\otimes 2}$.
In this subsection, we will introduce the purity estimation methods.

\begin{figure}[b]
    \centering
    \includegraphics[width=0.35\textwidth]{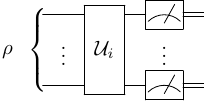}
    \caption{The circuit of random measurement.
        The gate $\mathcal{U}_i$ is uniformly sampled from an ensemble of unitary gates, like the n-qubit or tensor products of single-qubit Clifford unitaries.
        The quantum state $\rho$ can be recovered from the n-bit measurement outcome and the choice of the unitary gate $\mathcal{U}_i$}
    \label{fig: RM}
\end{figure}

The most trivial way to measure the purity is to perform the quantum state tomography (QST), and then calculate the purity by definition.
However, QST is so expensive that it is hard to perform on a large system.
Therefore, we need more efficient methods to measure purity.
The purity can be measured by random measurement~\cite{huang2020predicting}, where we randomly measure the operators sampled from an ensemble, typically the Pauli string of the qubits.
The circuit is shown in Figure~\ref{fig: RM}.
This method is called classical shadow, which can substitute QST.
By random measurement of operators with the Monte Carlo sampling, this method is more feasible than QST for large-size systems.
The number of measurements is shown in the order of
\begin{equation}
    \sim \frac{\log(M) \sup_i \Vert O_i\Vert}{\epsilon^2},
\end{equation}
which is determined by the observable operators $\hat{O}_i, (i=1,\dots, M)$ of interest, the ensemble of unitary gates, and the precision $\epsilon$.
However, since it is equivalent to QST in probability, it cannot solve the problem of exponential increasing dimension of the quantum system.

\begin{figure}[t]
    \centering
    \includegraphics[width=0.9\textwidth]{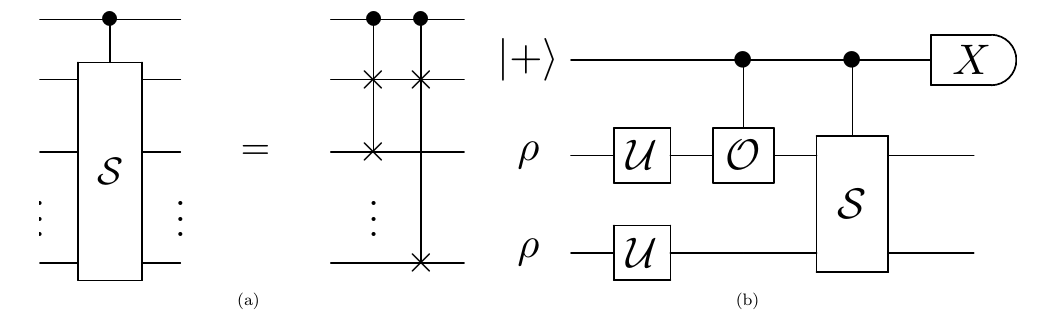}
    \caption{The circuit of (a) $\hat{S}^{(M)}$ and (b) the Hadamard test of VD/ESD.}
    \label{fig: purification}
\end{figure}

In the purification method, the measurement of quantities $\mathrm{Tr}(\rho^M)$ is the key problem.
The measurement method with a replica is proposed, which is founded on the following equation
\begin{equation} \label{eq: S}
    \mathrm{Tr}\left(\hat{O} \rho^M\right) = \mathrm{Tr}\left(\hat{O}_{(i)} \hat{S}^{(M)} \rho^{\otimes M}\right),
\end{equation}
where $\rho^{\otimes M}$ is $M$ copies of $\rho$, $\hat{O}_{(i)}$ is the operator $\hat{O}$ acted on the $i$-th copy, typically $i=1$, and $\hat{S}^{(M)}$ is cyclic permutation on $M$ copies
\begin{equation}
    \hat{S}^{(M)} \ket{\psi_1} \otimes \ket{\psi_2} \otimes \dots \otimes \ket{\psi_M} = \ket{\psi_2} \otimes \dots \otimes \ket{\psi_M} \otimes \ket{\psi_1}.
\end{equation}
The purity can be mapped to the expectation of $\hat{S}^{(2)}$ on two-replica state $\rho^{\otimes 2}$.
The circuit of $\hat{S}^{(M)}$ is shown in Figure~\ref{fig: purification}(a), and the circuit to measure the expectation in Equation~(\ref{eq: S}) is shown in Figure~\ref{fig: purification}(b).
This method allows us to measure the purity directly from the quantum circuit, which is generally used in the so-called virtual distillation or error suppression by derangement (VD/ESD) in the purification method~\cite{PhysRevX.11.041036}.
Although it can reduce the exponentially increasing quantities measured in QST, it enlarges the space complexity of qubits for the preparation of $2$ copies, and the SWAP gates needed in applying $\hat{S}^{(2)}$ will enlarge the time complexity of the quantum circuit.

\subsection{Data Processing} 

With the data $\braket{\hat{P}_i}_n$, $p_n$ from experiments, we calculate the effective error rate $s_n$ and the estimators $\braket{\hat{P}_i}_{n\mathrm{est}}$ in Equation~(\ref{eq: estimation}).
Then, we extrapolate the estimators to noise-free estimator $\braket{\hat{P}_i}_{\mathrm{est}}$, where $n = \frac{1}{2}$ or $s_{n_0} = 0$. 
Finally, calculate the estimator of ideal expectation $\braket{\hat{O}}_0 = \sum_i O_i \braket{\hat{P}_i}_{\mathrm{est}}$. 

Since the errors are Pauli diagonal with the Pauli twirling technique, as shown in Equation~(\ref{eq: expectation}) and~(\ref{eq: purity}), $\braket{\hat{P}_i}_{n}$ is an exponential function of $n$, while $p_n$ is a multi-exponential function of $n$, so the fitting model is 
\begin{align}
    \braket{\hat{P}_i}_{n} & = g(n) = B e^{-k n}, \\
    p_n & = f(n) = \sum_{i} A_i e^{- k_i n} + C.
\end{align}
Moreover, it can be simplified as 
\begin{equation}
    p_n = F(\braket{\hat{P}_i}_{n}) = \sum_i A_i \braket{\hat{P}_i}_{n}^{k_i} + C,
\end{equation}
where $(A_i, k_i, C)$ is the parameters need to determine.

In particular, the expectation $\braket{\hat{O}}_0$ can be calculated with the estimator $\braket{\hat{P}_i}_{1\mathrm{est}}$ without unitary folding and extrapolation. 
This should be interpreted as a modified VD/ESD of the purification method (see Appendix~\ref{app: purification}).   
In the following, we will discuss the difference between the two criteria of noisy-free point, and the difference between the $1$-fold pZNE (without extrapolation) and VD/ESD.

For extrapolation to $n = \frac{1}{2}$, $\chi_{ni} = \sqrt{\frac{\chi_{f_i}}{\chi_{bi}}} = e^{-\frac{\lambda^2}{2} \omega_i}$ by Equation~(\ref{eq: second}).
The tolerant error is
\begin{equation}
    \epsilon \sim \sqrt{2} \sigma \sim \frac{\sqrt{2}\lambda^2}{2} \Delta \bar{\omega} + O(\lambda^4),
\end{equation}
where $\Delta \bar{\omega} = \sqrt{\overline{\omega^2} - \bar{\omega}^2}$, has lower order than the bias of estimator $\braket{\hat{P}_i}_{1\mathrm{est}}$ without extrapolation.
For extrapolation to pure state, $s_{n_0} = 0$, the tolerant error is (see Appendix.~\ref{app: effectiveness})
\begin{equation}
    \epsilon \sim \sqrt{2} \sigma  \sim \sqrt{2} (1 -n_0) \lambda^2 \Delta \bar{\omega} + O(\lambda^4),
\end{equation}
which is in the same order as extrapolate to $n = {1}/{2}$, and in practice, we can select the one in $n_0$ and $1/2$ closer to $1$.

For routine ZNE method, the relative bias is $\left|e^{- \lambda^2 \omega_i/2} - 1\right|$.
The tolerant error is (see Appendix.~\ref{app: effectiveness}) 
\begin{equation}
    \epsilon \sim \frac{\sqrt{2}}{2} \lambda^2 \Delta \bar{\omega} + O(\lambda^4),
\end{equation}
when $\lambda^2 \Delta \bar{\omega} \leq \frac{1}{\sqrt{2} (1-n_0)} \sim \sqrt{2}$, otherwise it may fail.
The relative bias up to the lowest order is the same as the pZNE extrapolating to $n = \frac{1}{2}$, however, considering the constraint of pZNE $\sigma = \frac{\lambda^2}{2}\Delta \bar{\omega} \leq 2$ or $\lambda^2 \Delta \bar{\omega} \leq 4$, we find that the pZNE can be used in the region where the error rate is large so that the routine ZNE method may fail. 
The suppression of bias to second order $\lambda^2$ is from the assumption that the difference of forward and backward error is in the second order, which comes from the pulse-inverse technique~\cite{henao2023adaptive}. 
Without the pulse-inverse technique, the error rate $\lambda$ in the above calculation is substituted by $\lambda^{1/2}$. 

Then, we consider the VD/ESD method with $2$-replica (see Appendix~\ref{app: purification}).
The estimator of VD/ESD is 
\begin{equation}
    \braket{\hat{P}_i}_p = \frac{\braket{\hat{P}_i}_n}{p_n},
\end{equation} 
where $n = 2$.
Under the Pauli diagonal error, with Equation~(\ref{eq: expectation}) and Equation~(\ref{eq: purity}), the bias of the estimator is 
\begin{equation} \label{eq: VD/ESD_bias}
    \left|\frac{D \chi_{ni}}{(D - 1) \overline{\chi_n^2} + 1} - 1\right| \sim \left|\frac{D}{(D - 1) \bar{\chi}_n + 1/\bar{\chi}_n} - 1\right|
\end{equation}
which approaches to zero when $\bar{\chi}_n = 1$, the noise-free point, or $\bar{\chi}_n = \frac{1}{D - 1}$ if $\Delta \bar{\chi}_n \ll 1$.
Therefore, it is only reliable when error is small enough.
As the estimation in Appendix~\ref{app: effectiveness}, the tolerant error 
\begin{equation}
    \epsilon \sim \sqrt[4]{12} \lambda > \sqrt{2} \lambda
\end{equation}
is larger than the one for the $1$-fold pZNE method.
Notice that the requirement $\Delta \bar{\chi}_n \ll 1$ of the $1$-fold pZNE method is weaker than that $\lambda \ll 1$ of VD/ESD.
Therefore, the $1$-folded pZNE has lower tolerant error and looser conditions than the VD/ESD method with $2$-replica.
\subsection{Efficiency} \label{subsec: efficiency}

To describe the efficiency of this method, we consider its overhead~\cite{RevModPhys.95.045005,cai2021practical}.
The overhead depends on the fitting model,  and here, we abstractly denote the fitting model as
\begin{equation}
    O_{em} = O(0) = F(\{p_i, O_i\}),
\end{equation}
where $F(\{p_i, O_i\})$ is a function of the data $\{p_i, O_i\}$ of purity and observable expectation from the experiment.
Then, we calculate the overhead $C_{em}$ of pZNE, and the details of calculation is in Appendix~\ref{app: efficiency}.

According to the law of large numbers (LLN), the overhead is given by 
\begin{align} 
    C_{em} & \equiv \frac{N^{\epsilon}(O_{em})}{N^{\epsilon}(O)} = \frac{\mathrm{Var} [O_{em}]}{\mathrm{Var}[O]} \nonumber\\
    & \lesssim \left(\sqrt{C_{em}^{\textrm{ZNE}}} + \frac{\mathrm{Var}[p]}{\mathrm{Var}[O]} \sum_i \left|\frac{\partial F}{\partial p_i}\right|\right)^2,
\end{align}
where $N^{\epsilon}(O_{em})$ and $N^{\epsilon}(O)$ is the number of shots to measure the mitigated estimator $O_{em}$ and the raw estimator $O$ up to $\epsilon$ precision correspondingly, and the overhead of routine ZNE method $C_{em}^{\textrm{ZNE}} \sim \sum_i \left({\partial F}/{\partial O_i}\right)^2$.
The variance $\mathrm{Var}[p]$ depends on the method to measure the purity, and as we previously discussed, there are two kinds of methods to measure the purity.
One is the replica measurement method, which is based on the Equation~(\ref{eq: S}).
Another one is the QST-like method, QST and classical shadow, which measure the expectation of all Pauli operators $\braket{\hat{\sigma}_{\alpha}}$, where $\alpha \in \{1, 2, \dots, 4^{L}-1\}$, and $L$ is the system size.
Then, we calculate the overhead of these two different measurement methods respectively.
(For details, see Appendix~\ref{app: efficiency}.)

For the replica measurement method, if we assume the observable of interest is a Pauli operator, and when the error rate $x \rightarrow 1$, $O \not \rightarrow 1$, then the overhead $C_{em}$ is 
\begin{equation}
    C_{em} \sim C_{em}^{\textrm{ZNE}}.
\end{equation}
In the contrary, if $O \rightarrow 1$, the overhead increases by a constant 
\begin{equation}
    C_{em} \lesssim \left(\sqrt{C_{em}^{\textrm{ZNE}}} + 2 \kappa \sum_i \left|\frac{\partial F}{\partial p_i}\right|\right)^2.
\end{equation}

For the QST-like measurement method, we assume the observable $\hat{O}$ is a Pauli operator located on $l_0$ qubits, namely $\hat{O}$ is not identity $\hat{I}$ exactly on $l_0$ qubits, and we call it $l_0$-weight Pauli operator in below.
We measure the $3^L$ different $L$-weight Pauli operators with the same number $N_L$ of shots to perform the QST-like method.
The overhead is
\begin{equation}
    C_{em} \lesssim \left(\sqrt{\tilde{C}_{em}^{\textrm{ZNE}}} + \frac{10^L -1}{4^{L-1} 3^{l_0}} \sum_i \left|\frac{\partial F}{\partial p_i}\right|\right)^2.
\end{equation}
If $l_0 > L \log_{3}\frac{5}{2}$, when $L \rightarrow \infty$, the second term in the rightest inequality goes to zero, which means that the overhead
\begin{equation}
    \lim_{L \rightarrow \infty} C_{em} \sim \tilde{C}_{em}^{\textrm{ZNE}}.
\end{equation}
Thus, the variance of purity $\mathrm{Var}[p]$ is well controlled by the variance of Pauli operator $\mathrm{Var}[O]$ with large weight.

However, we note that the overhead of ZNE $\tilde{C}_{em}^{\textrm{ZNE}}$ here is calculated under the assumption that the expectation of $\braket{\hat{O}}$ is measured from QST-like method.
In the ZNE method, we need not perform QST to measure $\braket{\hat{O}}$, thus the actual ZNE overhead is
\begin{equation}
    C_{em}^{\textrm{ZNE}} \sim 3^{-L} Z \tilde{C}_{em}^{\textrm{ZNE}},
\end{equation}
where $Z$ is the number of different $L$-weight operators covering the operator $\hat{O}_i$ of interest.
Therefore, the overhead asymptotically is
\begin{equation}
    \lim_{L \rightarrow \infty} C_{em} \lesssim \frac{3^{L}}{Z} C_{em}^{\textrm{ZNE}}.
\end{equation}
Moreover, if the operators of interest is so dense that every $L$-weight operator covers at least one operator $\hat{O}_i$ of interest, $Z = 3^L$, the efficiency of pZNE is of the same order as the routine ZNE.  

\begin{figure}[t]
    \centering
    \includegraphics[width=0.85\textwidth]{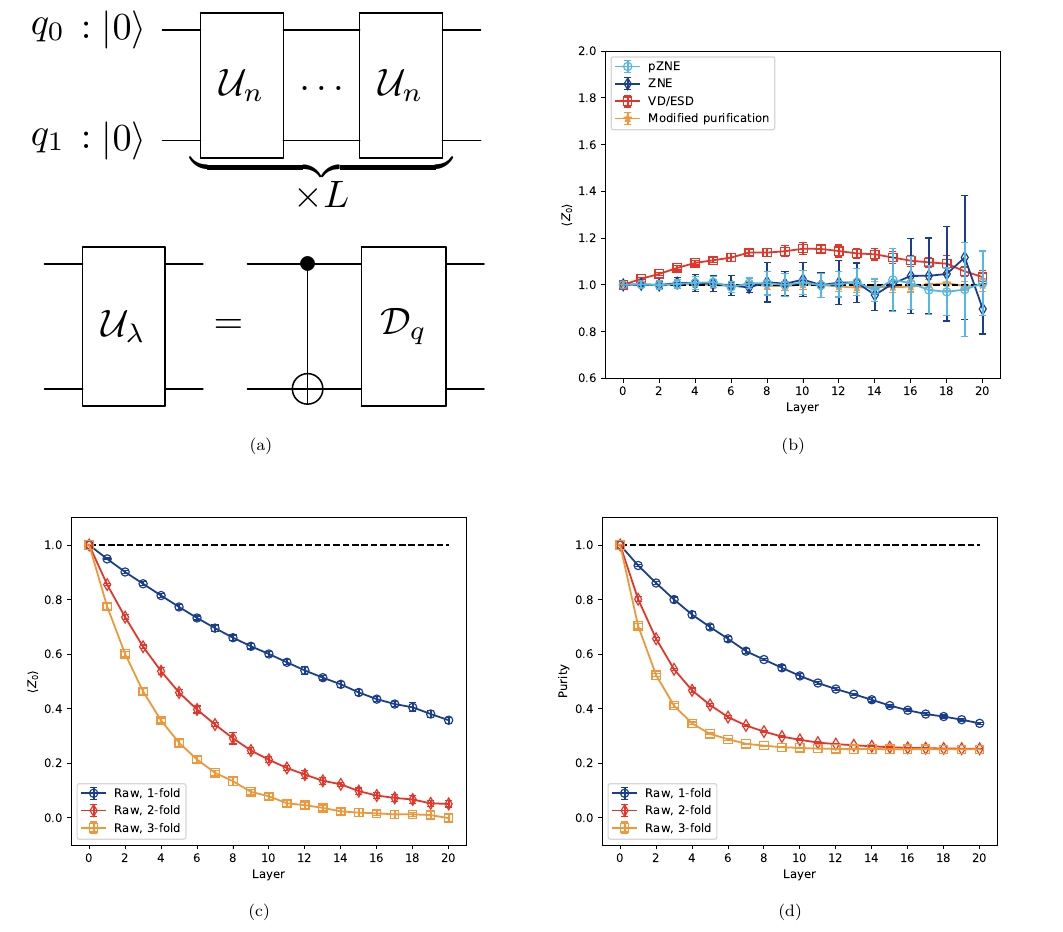}

    \caption{
        The numerical simulation of pZNE, ZNE, and VD/ESD with the depolarizing error $\mathcal{D}_q$, where error rate $q= 0.05$ per gate.
        The circuit is shown in~(a), which includes several layers of folded noisy CNOT gate $\mathcal{U}_n$.
        Each circuit is simulated with $10$ repetitions, and the QST results are obtained by performing $1000$ single-shot measurements on each $2$-weight Pauli basis.
        The mitigated expectations of pZNE, ZNE, and VD/ESD are shown in~(b).
        The raw data of expectation $\braket{\hat{Z}_0}$ and purity are shown in~(c) and~(d). 
    }
    \label{fig: run_sim}
\end{figure}

\begin{figure}[t]
    \centering
    \includegraphics[width=0.85\textwidth]{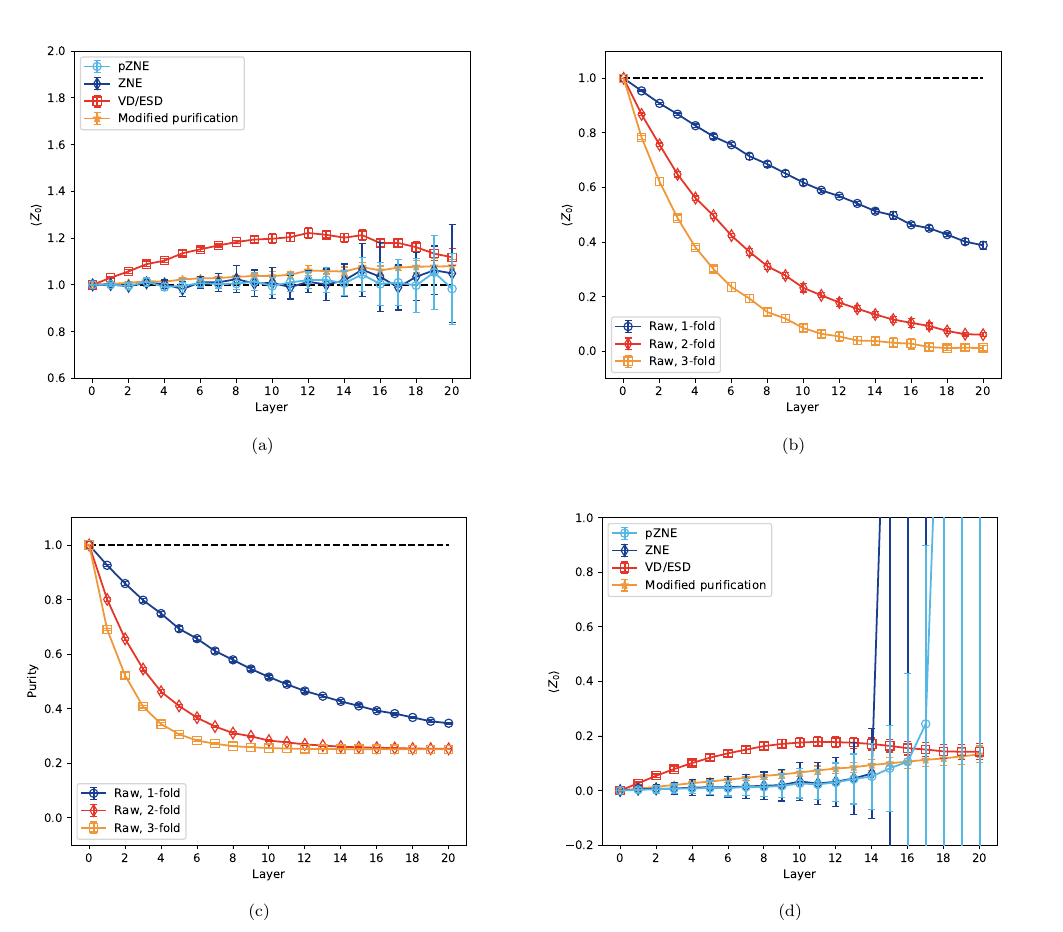}

    \caption{
        The numerical simulation of pZNE, ZNE, and VD/ESD with the randomly sampled Pauli diagonal error $\mathcal{D}_q$ with fixed error probability $q = 0.05$ per gate. 
        The circuit is the same as Figure~\ref{fig: run_sim} under the Pauli diagonal error $\mathcal{D}_q$ selected.
        Each circuit is simulated with $10$ repetitions, and the QST results are obtained by performing $2000$ single-shot measurements on each $2$-weight Pauli basis.
        The mitigated expectations of pZNE, ZNE, and VD/ESD are shown in~(a), and the raw data of expectation $\braket{\hat{Z}_0}$ and purity in~(b) and~(c). 
        The results of average over $50$ random sampled Pauli diagonal errors are shown in~(d).
        The data points show the value of $\Delta \braket{Z_0}_1$ and error bars show the value of $\Delta \braket{Z_0}_2$.
    }
    \label{fig: run_pauli}
\end{figure}

\begin{figure}[t]
    \centering
    \includegraphics[width=0.85\textwidth]{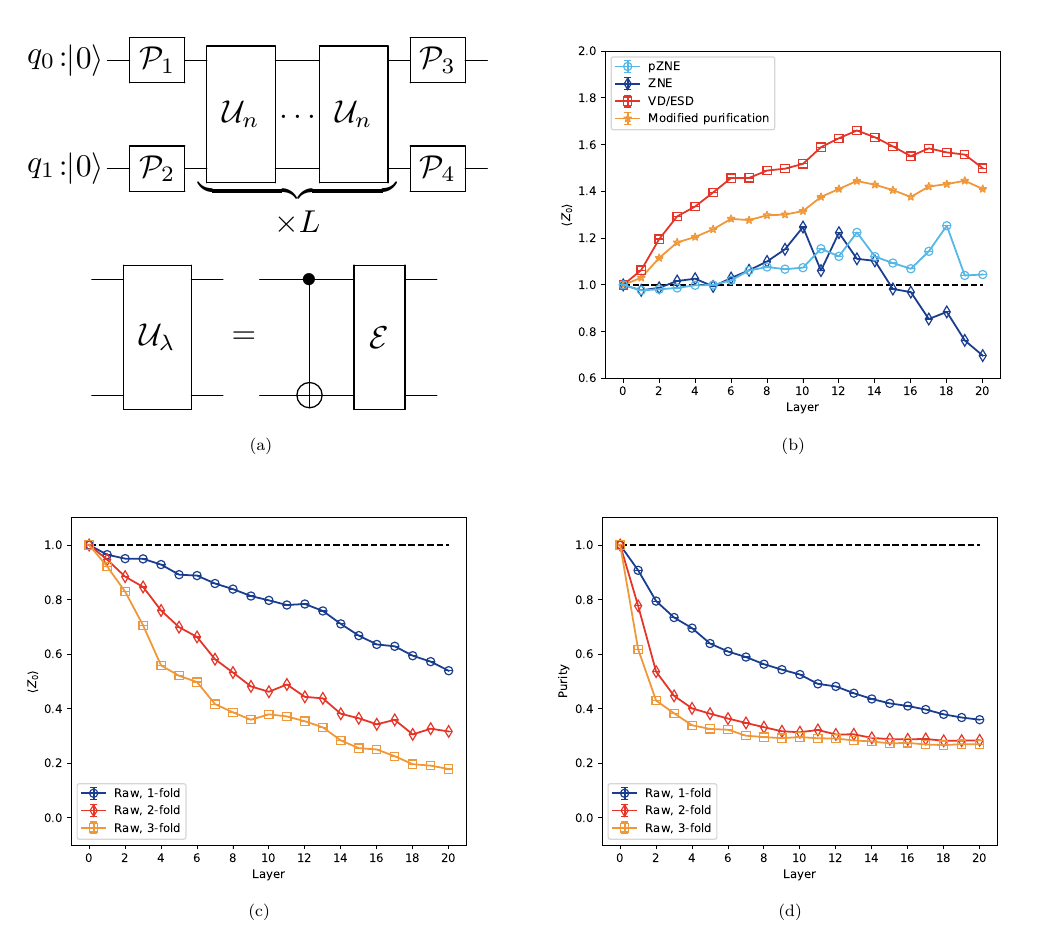}
    \caption{
        The experimental simulation of pZNE, ZNE, and VD/ESD.
        The Pauli twirled circuit is shown in~(a), which consists of layers of folded noisy CNOT gate $\mathcal{U}_n$, where the noise $\mathcal{E}$ is determined by the experiment device.
        The expectation $\braket{\hat{Z}_0}$ with the initial state $\ket{00}$ under increasing CNOT gate folding layers is shown.
        The fitting model is an exponential model.
        Each Pauli basis is repeatedly measured with $1600$ shots, whereas each Pauli twirling instance is performed with $100$ shots.
        The mitigated expectations of pZNE, ZNE, and VD/ESD are shown in~(b).
        The raw data of expectation $\braket{\hat{Z}_0}$ and purity are shown in~(c) and~(d). 
    }
    \label{fig: exper}
\end{figure}

\section{Verification} \label{sec: simu}

In this section, we simulate the pZNE method numerically and experimentally.

\subsection{Numerical Simulation: Depolarizing Error}

We verify the effectiveness of pZNE by numerical simulation.
First, we simulate the CNOT gates under the depolarizing error of different error rates.
The initial state is selected as $\ket{00}$, and the observable expectation $\braket{\hat{Z}_0}$ of Pauli $Z$ operator on the qubit $q_0$ is measured for showing the effect of the error.
For ideal CNOT gates, the expectation $\braket{\hat{Z}_0} = 1$ is invariant for any number of folds.
However, the depolarizing error will lead to the decay of the expectation.
We compare the mitigated expectation of the pZNE to the routine ZNE method, the VD/ESD method, and the modified purification method.
Each layer of CNOT gate is a folded gate, with $1,3$ and $5$ CNOT gates.
We simulate the circuit with different layers, which enlarge the error rate of $1$-fold gate.
The results are shown in Figure~\ref{fig: run_sim}.
The estimators of pZNE, ZNE, VD/ESD, and modified purification method are shown in Figure~\ref{fig: run_sim}(b), the raw value of expectations $\braket{\hat{Z}_0}$ are shown in the Figure~\ref{fig: run_sim}(c), and the purities of the noisy state are shown in the Figure~\ref{fig: run_sim}(d). (For details of data processing, see Appendix~\ref{app: simulation}.)

Among the several methods, the VD/ESD method presents a large bias of expectation, which is expected in Equation~(\ref{eq: VD/ESD_bias}).
The modified purification has the lowest bias and standard deviation since the depolarizing error $\Delta \chi = 0$. 
However, in cases where the error is not a depolarizing error but a Pauli diagonal error, the modified purification method may exhibit a significant bias that increases with the error rate.
We will discuss this later in Sec~\ref{subsec: pauli_error}.

The reason pZNE (with extrapolation) does not perform better than the modified purification, $1$-fold pZNE without extrapolation, is that for the depolarizing errors, the error is from the random errors in measurement.
The extrapolation thus cannot cancel the bias, but enlarge the random errors in measurement. 
We expect that there exist errors with small but not vanish $\Delta \chi$, where the modified purification method perform better than the routine ZNE method.
With sufficiently large number $N$ of shots, the random errors in measurement are sufficiently smaller than the error from error models.
In this case, the pZNE (with extrapolation) will perform better than both the modified purification and the routine ZNE method. 

Comparing the pZNE with the ZNE, the bias between the average mitigated data and ideal expectation of these two methods are comparable before the $14$-th layer, where the purity of $1$-folded noisy state $p \geq 0.5$, as shown in Figure~\ref{fig: run_sim}(d).
It can be seen that the standard deviations of the pZNE method are lower than those of the ZNE method. Despite this, purity estimation necessitates complete information about the quantum state, resulting in an additional measurement overhead.
However, in scenarios where dense information about the state is required, the purity can be estimated simultaneously from the measurement outcomes. In such cases, pZNE does not involve higher overhead compared to ZNE, and it showcases the advantage of offering higher precision than ZNE.



\subsection{Numerical Simulation: Pauli Diagonal Error} \label{subsec: pauli_error}
    
    To analyze the pZNE method with the Pauli diagonal error, we first simulate the CNOT gate under this kind of error models. The settings of simulations are the same as those in the depolarizing error model case.
    The results are shown in Figure~\ref{fig: run_pauli}. 
    The estimators of pZNE, ZNE, VD/ESD, and modified purification method of $1$-folded gate are shown in Figure~\ref{fig: run_pauli}(a), the raw value of expectations $\braket{\hat{Z}_0}$ are shown in Figure~\ref{fig: run_pauli}(b), and the purities of the noisy state are shown in Figure~\ref{fig: run_pauli}(c). (For details, see Appendix~\ref{app: simulation}.)
    
    The VD/ESD estimator still has a large bias from the ideal expectation.
    Different from the depolarizing error model case, the modified purification method also has bias with randomly selected Pauli diagonal error.
    The bias increases with the increasing of the layer in Figure~\ref{fig: run_pauli}(a), which represents the increasing error rate.
    This result is because the eigenvalues of the selected error channel do not concentrate enough.
    For the pZNE and routine ZNE method, same as the depolarizing error case, the bias of pZNE is comparable with the routine ZNE.
    In addition, the standard deviation of pZNE is lower than routine ZNE.

    However, the observation from the results of the depolarizing error and the selected Pauli diagonal error are not sufficient to represent the performance of different methods.
    To confirm this, we perform the numerical simulations of the CNOT gate under a wider range of randomly sampled Pauli diagonal errors with the same error rates. We take 50 instances of Pauli diagonal errors, randomly sampled, and computed the expectations and variances of various mitigation methods for each instance. Subsequently, we calculated the square root of the second moment of biases for these sampled instances 
    \begin{equation}
        \Delta\braket{Z_0}_1 = \left[\mathbb{E}_{\mathcal{E}_i} (\braket{Z_0} - 1)^2\right]^{1/2},
    \end{equation}
    and the average standard deviation
    \begin{equation}
        \Delta \braket{Z_0}_2 = \left[\mathbb{E}_{\mathcal{E}_i}(\mathrm{Var}\braket{Z_0})\right]^{1/2},
    \end{equation}
    to represent the average performance of the different methods.

    The quantity $\Delta\braket{Z_0}_1$ represents the accuracy of the methods, and the quantity $\Delta \braket{Z_0}_2$ represents the precision.
    The result is shown in Figure~\ref{fig: run_pauli}(d).

    The VD/ESD method exhibits a noticeable average bias, whereas the bias of the modified purification method increases linearly with the number of layers. Interestingly, their standard deviations remain comparable and do not show an increase in the number of layers.
    In comparison, the biases of the pZNE and conventional ZNE methods are nearly equivalent and tend to increase with the number of layers. However, before the $14$-th layer, the bias of pZNE is marginally smaller than that of conventional ZNE. After the $15$-th layer, conventional ZNE encounters failures for some sampled instances due to their diverging biases, while pZNE manages to remain effective until the $18$-th layer.
    Furthermore, the standard deviations of pZNE show a corresponding increase with the number of layers, as both methods involve extrapolation. Notably, these standard deviations are consistently lower than those of the routine ZNE method.

    These results show that the pZNE method offers enhanced accuracy and precision, particularly in scenarios with substantial error rates, while also maintaining a lower standard deviation compared to routine ZNE on average across various Pauli diagonal errors. Nonetheless, to harness these benefits effectively, it is recommended to leverage the pZNE method primarily in tasks that necessitate detailed and comprehensive information about the quantum state.

\subsection{Experiment}
    We perform the pZNE and ZNE experiments on the cloud-based superconducting quantum computation platform, \emph{Quafu}~\cite{chen2022scq}. For the information of the Quafu platform and device, see Appendix~\ref{app: experiment}. The experimental setup was identical to that of the numerical simulations but was performed only once. We applied the Pauli twirling technique to the folded CNOT gate. The results are illustrated in Figure~\ref{fig: exper}, where the mitigated outcomes of pZNE, ZNE, VD/ESD, and the modified purification method of the $1$-folded gate are presented in Figure~\ref{fig: exper}(b). Additionally, the raw expectation values $\braket{\hat{Z}_0}$ are shown in Figure~\ref{fig: exper}(c), and the purities of the noisy state are shown in Figure~\ref{fig: exper}(d). The data processing procedures were the same as those used in the numerical simulation, with the exception of calculating standard deviations.
    
    The results of the VD/ESD and modified purification methods exhibit significant biases away from the ideal values, indicating a lack of concentration in the eigenvalues of the twirled error channel. The mitigated expectations of the pZNE show biases comparable to those of the standard ZNE method. Interestingly, for error rates in the moderate range, specifically after the $5$-th layer and before the $11$-th layer, the bias of the pZNE method is lower than that of the standard ZNE approach.
    Our previous assumption considered each error channel of the layers in the circuit to be Pauli diagonal. However, implementing individual Pauli twirling for each layer in the circuit would lead to an exponential increase in compiled circuit instances with the number of layers. To simplify this process, we twirled the folding circuit as a whole. Consequently, the coherent part of the error channels, which would have been eliminated in individual Pauli twirling, contributed to the overall error channel of longer layers. 
    In such cases, the purity of the state can assist in extrapolating expectations, thereby reducing the bias in pZNE expectations compared to routine ZNE for stages with moderate errors.

\section{Conclusion} \label{sec: conclusion}
In this paper, we introduce a method that utilizes purity to assess noise error rates, operating under the assumption that the error channel is Pauli diagonal. By using the Pauli twirling technique, we ensure that this assumption holds universally. Our study involves estimating the bias of the pZNE method based on a specified failing probability and conducting a comparative analysis with other relevant methods. We demonstrate that the pZNE method may not lead to a substantial reduction in bias compared to the routine ZNE method when error rates are low. However, it showcases its utility within a certain threshold where the routine ZNE method might encounter limitations or failures.

Moreover, we analyze the overhead of the pZNE method using different purity estimation techniques, such as the replica measurement method and the QST-like method. Our results reveal distinct behaviors in the upper bounds of pZNE for these estimation methods. Specifically, the overhead of the replica measurement method is constrained by the overhead of ZNE with a finite constant added, while the overhead of a QST-like method escalates exponentially with the growth of system size. In cases with exponentially increasing overhead, the pZNE method may not be efficient for large systems when using a QST-like method. However, if the task requires dense information about the density matrix, the purity estimation will not significantly inflate the overhead. 

In addition, we verify the pZNE method with the numerical simulations under the depolarizing error and the Pauli diagonal errors, as well as the experiments on the quantum computation platform, \emph{Quafu}. We compare it with the routine ZNE and the VD/ESD method. The results show that the bias of both routine ZNE and pZNE increases with the error rate, and divergence when the error rate is large enough. The bias of pZNE is slightly lower than routine ZNE when the error rate is moderate, and the critical point of error rate for pZNE is larger than routine ZNE method. The standard deviation of pZNE is lower than that of ZNE, indicating that the pZNE method is more efficient when dense state information is of interest.

Our results show that the purity of noisy quantum states can be used to assist the ZNE method, thus extending its effectiveness.
It can be expected that pZNE will reinforce the wide usage of the ZNE method in the NISQ era.

\appendix
\section{The Forward and Backward Errors} \label{app: errors}

    In this appendix, we consider the forward and backward errors.
    The equality of forward and backward error is the basic assumption of the ZNE method.
    However, it is not such easy to satisfy, even for the unitary evolution whose inverse is itself.
    For example, assume the ideal unitary $\mathcal{U}$ is CNOT gate, the noisy circuit is $\mathcal{U}_{\lambda} = \mathcal{E}_f \circ \mathcal{U}$.
    Since the CNOT gate is self-inverse, in the unitary folding technique, the backward evolution is $\mathcal{U}_{\lambda}^{\dagger} = \mathcal{U}_{\lambda}$, and the backward error is 
    \begin{equation}
        \mathcal{E}_b = \mathcal{U} \circ \mathcal{E}_f \circ \mathcal{U}^{\dagger}.
    \end{equation}
    We assume the circuit is implemented with the Pauli twirling technique, thus the forward and backward errors are Pauli diagonal.
    Let the Pauli operators on the first control qubit of CNOT gate be $X_1, Y_1, Z_1$, and on the target qubit be $X_2, Y_2, Z_2$.
    The action of CNOT gate is 
    \begin{align} \label{eq: propogation}
        X_1 \mapsto X_1 X_2, \quad Y_1 \mapsto Y_1 X_2, \quad Z_1 \mapsto Z_1, \nonumber \\
        X_2 \mapsto X_2, \quad Y_2 \mapsto Z_1 Y_2, \quad Z_2 \mapsto Z_1 Z_2,
    \end{align}
    so if there is no symmetry that all coefficients $q_i$ of Pauli diagonal forward error $\mathcal{E}_f$ are different (the converse case is zero-measure in space of error), then the forward and backward error is different.

    In Ref.~\cite{henao2023adaptive}, they deal with this problem by introducing the pulse-inverse circuit $\mathcal{U}_{\lambda I}^{\dagger}$ ($\mathcal{K}_I$ in original context).
    The pulse-inverse circuit $\mathcal{U}_{\lambda I}^{\dagger}$ is a special circuit of inverse of $\mathcal{U}_{\lambda}$.
    Assume the ideal evolution $\mathcal{U}$ is implemented by a time-dependent Hamiltonian $H(t)$, and the error is described by a Lindblad operation $\mathcal{L}(t)$.
    The ideal pulse-inverse circuit $\mathcal{U}^{\dagger}$ is implemented by time-dependent Hamiltonian $H_I(t) = - H(T-t)$, and a reasonable assumption is that the error in this inverse pulse is described by the Lindblad operation $\mathcal{L}_I(t) = \mathcal{L}(T - t)$.
    The forward and backward error thus is related to the evolution in the interaction picture
    \begin{align}
        \mathcal{E}_f & = \mathcal{U} \circ \mathrm{T} \exp \int_0^T\mathrm{d}t \tilde{L}(t) \circ \mathcal{U}^{\dagger},  \\
        \mathcal{E}_b & = \mathrm{T} \exp \int_0^T\mathrm{d}t \tilde{L}_{I}(t),  \\		
    \end{align}
    where $\tilde{L}(t) = \mathcal{U}^{\dagger}(t) L(t) \mathcal{U}(t), \tilde{L}_{I}(t) = \mathcal{U}_I^{\dagger}(t) L_I(t) \mathcal{U}_I(t)$ are the Lindblad operations in the interaction picture.
    Ref.~\cite{henao2023adaptive} shows that in the first order of Magnus expansion $\mathcal{E}_f \approx \mathcal{E}_b$, which agrees with the first order expansion of time-independent case.  
    However, to higher order, these two errors are still different, if the Hamiltonian evolution and Lindblad evolution are not commute.
    Therefore, the pulse-inverse circuit extends the requirement of time-independence on unitary folding to the Floquet system, which is more suitable for digital systems, but cannot fix the problem in the assumption that the forward and backward errors are equal. 

    Moreover, intersecting a perturbative parameter $\lambda$ before Lindblad operations to reflect the intensity of the noise, the expansion shows that 
    \begin{equation}
        \mathcal{E}_b = \mathcal{E}_f e^{\lambda^2 \Omega_2},
    \end{equation}
    the difference between forward and backward errors is up to the second order.
    With the Pauli twirling technique, both forward and backward errors are Pauli diagonal, thus commute, so their eigenvalues have a relation 
    \begin{equation} \label{eq: second}
        \chi_{bi} = \chi_{fi} e^{\lambda^2 \omega_i}.
    \end{equation}

\section{Brief Introduction to Purification Method} \label{app: purification}

The purification method is based on the belief that in practice the output state is always mixed, while the ideal state is pure.
If the error is not so large, the ideal pure state should not be far away from the experimental mixed state.
Thus, the pure state closest to the mixed state may be a good approximation of the ideal state, and this state is the aim of the purification method.

Let the mixed state $\rho$ decomposed as 
\begin{equation}
    \rho = \sum_{n} p_n \ket{n} \bra{n},
\end{equation}
where $p_n$ are the spectrum of the density ordered descendingly, and $\ket{n}$ are the corresponding eigenstates.
For some other pure state $\ket{\psi} = \sum_{n} \psi_n \ket{n}$, the distance
\begin{equation}
    \begin{split}
        D^2(\rho, \ket{\psi}) & = \mathrm{Tr} \rho^2 + 1 - 2 \bra{\psi} \rho \ket{\psi} \\
        & = \mathrm{Tr} \rho^2 + 1 - 2 \sum_n p_n |\psi_n|^2    
    \end{split}
\end{equation}
minimizes when $|\psi_n| = \delta_{n,0}$, namely the state with the largest probability in the mixed state is the closest pure state.
It is clear that if we know the mixed state completely, by diagonalization we can get the closest pure state.

For large systems, the dimension of the matrix is too overwhelming to perform the diagonalization.
The better way is to realize the purification by iteration.
One of the iteration procedures is the McWeeny purification \cite{RevModPhys.32.335,google2020hartree}
\begin{equation}
    \rho_{n+1} = 3 \rho_n^2 - 2 \rho_n^3.
\end{equation}
In a basis that can diagonalize the mixed state, this iteration only maps the spectrum.
However, the attractive fixed point of this map is $p = 0, 1$ and the critical point is $1/2$, thus it has the probability of convergence to $0$ for all eigenvalues, which is not what we desire.
Only if there is an eigenvalue $p_n$ greater than $1/2$, namely the error is small enough, the purification is working.
Another one is the map 
\begin{equation}
    \rho \mapsto \frac{\rho^M}{\mathrm{Tr} \rho^M}.
\end{equation} 
In the diagonal representation, the spectra are mapped as
\begin{equation}
    p_n \mapsto \frac{p_n^M}{\sum_{m} p_m^M} = \frac{1}{1 + \sum_{m \neq n} (p_m/p_n)^M} .
\end{equation}
In the limit $M \rightarrow \infty$, the spectra $p_n \rightarrow \delta_{n 0}$, since $p_n$ is in descending order, which is what pure state we want.

To do the purification, one can perform the tomography of target density, and then iterate the procedure given above on a classical computer.
However, tomography is expensive for large systems, other method is also needed.
One can prepare the purified state by post-selection, Other strategy is to realize the purification virtually.
The purified state can be approximated as $\frac{\rho^M}{\mathrm{Tr} \rho^M}$ for some given integer $M$, so the expectation of quantity $\hat{O}$ under the purified state approximate as
\begin{equation}
    \braket{\hat{O}}_{\mathrm{p}} \approx \frac{\mathrm{Tr}\left(\hat{O} \rho_{\lambda}^M\right)}{\mathrm{Tr} \rho_{\lambda}^M}.
\end{equation} 
If the value $\mathrm{Tr}\left(\hat{O} \rho_{\lambda}^M\right),\mathrm{Tr} \rho_{\lambda}^M$ can be evaluated from the experiment, the purified expectation value can be obtained.

A method called virtual distillation or error suppression by derangement is proposed \cite{PhysRevX.11.041036}, which is based on the Equation~(\ref{eq: S}).
where $\rho^{\otimes M}$ is $M$ copies of $\rho$, $\hat{O}_{(i)}$ is the operator $\hat{O}$ acted on the $i$-th copy, typically $i=1$, and $\hat{S}^{(M)}$ is cyclic 
Thus, $\mathrm{Tr}\left(\hat{O} \rho_{\lambda}^M\right)$ and $\mathrm{Tr}\rho_{\lambda}^M$ can be measured with $M$ copies of $\rho$. 
For $\hat{S}^{(M)}$ is cyclic permutation on $M$ copies, the expectation of operator $\hat{O}^{(M)} = \frac{1}{M} \sum_i \hat{O}_{(i)}$ is same as the original one
\begin{equation}
    \mathrm{Tr}\left(\hat{O}_{(i)} \hat{S}^{(M)} \rho^{\otimes M}\right) = \mathrm{Tr}\left(\hat{O}^{(M)} \hat{S}^{(M)} \rho^{\otimes M}\right).
\end{equation}
Since $\hat{O}^{(M)}$ are commute with $\hat{S}^{(M)}$, we can find the common eigenstates as the basis, which allows measuring $\mathrm{Tr}\left(\hat{O} \rho^M\right)$ and $\mathrm{Tr}\rho^M$ simultaneously.
One can measure the expectation $\mathrm{Tr}\left(\hat{O}^{(M)} \hat{S}^{(M)} \rho^{\otimes M}\right)$ and $\mathrm{Tr}\left(\hat{S}^{(M)} \rho^{\otimes M}\right)$ by Hadamard test (see Figure~\ref{fig: purification}) or by diagonalization \cite{PhysRevX.11.041036}.
For the measurement by diagonalization, in a qubit system, let $\hat{O}$ be the $\hat{Z}$ operator, the eigenstates are just spin waves.
Moreover, considering the case with two copies, i.e. $M=2$, the eigenstates states are just the spin singlet and triplet, and the transformation is 
\begin{equation}
    \hat{B}^{(2)} = \left(\begin{array}{cccc}
        1 & 0 & 0 & 0 \\
        0 & \frac{1}{\sqrt{2}} & -\frac{1}{\sqrt{2}} & 0 \\
        0 & \frac{1}{\sqrt{2}} & \frac{1}{\sqrt{2}} & 0 \\
        0 & 0 & 0 & 1
    \end{array} \right).
\end{equation}
Under this transformation, 
\begin{equation}
    \begin{split}
        \hat{S}^{(2)} & \rightarrow \frac{1}{2}(1 + \hat{Z}_{(1)} - \hat{Z}_{(2)} + \hat{Z}_{(1)} \hat{Z}_{(2)}) \\
        \hat{Z}^{(2)} & \rightarrow \hat{Z}^{(2)} = \frac{1}{2} (\hat{Z}_{(1)} + \hat{Z}_{(2)}),
    \end{split}
\end{equation}           
and we can verify that $\hat{Z}^{(2)} \hat{S}^{(2)} = \hat{Z}^{(2)}$.
Thus, the estimator of $\hat{Z}$ is 
\begin{equation}
    \braket{\hat{Z}}_p \approx \frac{\braket{\hat{Z}}_{\lambda}}{\mathrm{Tr} \rho_{\lambda}^2}.
\end{equation} 

In addition to the above-mentioned methods, there is a variation of purity estimation called state verification~\cite{cai2021resourceefficient}.
We consider the purity of $\rho_n = \mathcal{U}_n(\rho)$, namely
\begin{align}
    p_n & = \mathrm{Tr} \rho_n^2 = \mathrm{Tr} \left[\mathcal{U}_n(\rho)^2\right] \nonumber\\
    & = \mathrm{Tr} \left[\rho \left(\mathcal{U}_n\right)^{\dagger} \circ \mathcal{U}_n(\rho)\right]  = \mathrm{Tr} \left[\rho \tilde{\rho}_n\right].
\end{align}
This is the overlap of initial state $\rho$ with quasi-state $\tilde{\rho}_n = \left(\mathcal{U}_n\right)^{\dagger} \circ \mathcal{U}_n(\rho)$, which is not physical in general.
Here, the $\dagger$ on the channel $\mathcal{U}_n$ is acted on its Kraus operators, but not on the output state $\mathcal{U}_n(\rho)$.
For $\mathcal{U}_n = (\mathcal{U}_{\lambda} \circ \mathcal{U}_{\lambda}^{\dagger})^{n-1} \circ \mathcal{U}_{\lambda}$, the quasi-channel is
\begin{equation}
    \left(\mathcal{U}_n\right)^{\dagger} \circ \mathcal{U}_n = \left(\mathcal{U}_{\lambda}\right)^{\dagger} \circ (\mathcal{U}_{\lambda} \circ \mathcal{U}_{\lambda}^{\dagger})^{\dagger n-1} \circ (\mathcal{U}_{\lambda} \circ \mathcal{U}_{\lambda}^{\dagger})^{n-1} \circ \mathcal{U}_{\lambda}.
\end{equation}
In ideal case, $\lambda = 0$, $\left(\mathcal{U}_{\lambda}\right)^{\dagger} = \mathcal{U}_{\lambda}^{\dagger} = \mathcal{U}^{\dagger}$, we have
\begin{equation}
    \left.\left(\mathcal{U}_n\right)^{\dagger} \circ \mathcal{U}_n \right|_{\lambda = 0} = \left.\tilde{\mathcal{U}}_{2 n - 1} \right|_{\lambda = 0},
\end{equation}
where $\tilde{\mathcal{U}}_n = (\mathcal{U}_{\lambda}^{\dagger} \circ \mathcal{U}_{\lambda})^n$.
Thus, in the purification method, the recurrence probability can be used instead of purity
\begin{equation}
    \tilde{p}_{n} = \mathrm{Tr} \left[\rho \rho_{n}\right],
\end{equation}
where $\rho_{n} = \tilde{\mathcal{U}}_{2 n - 1}(\rho)$ is a physical state, to substitute the purity.

The state verification echo $\tilde{p}_{n}$ is easier to measure than purity.
However, it is coincident with the purity if the error channels satisfy
\begin{equation}
    \left(\mathcal{E}_f\right)^{\dagger} = \mathcal{E}_b.
\end{equation}
In the Pauli twirling case, it is equivalent to $\mathcal{E}_f = \mathcal{E}_b$, where the routine ZNE method is satisfied.
Therefore, it cannot be used in pZNE to measure purity.

\section{Pauli Twirling} \label{app: pauli_twirling}

    Pauli twirling technique can compile non-diagonal error channel into Pauli diagonal.
    Given a group $G$ of gates the twirling channel of $\mathcal{E}$ with these gates is 
    \begin{equation}
        \mathcal{E}_{\mathrm{twirl}} = \frac{1}{|G|} \sum_{\mathcal{G} \in G} \mathcal{G} \circ \mathcal{E} \circ \mathcal{G}^{\dagger}.
    \end{equation}
    If the gate $G$ is Pauli group, writing the error channel $\mathcal{E}$ as the Kraus decomposition in Pauli basis
    \begin{equation}
        \mathcal{E} = \sum_i e_{ij} \hat{P}_i \cdot \hat{P}_j,
    \end{equation}
    the twirling channel is 
    \begin{align}
        \mathcal{E}_{\mathrm{twirl}} & = \frac{1}{D^2} \sum_{i,j, k} e_{ij} \mathcal{P}_{k}(\hat{P}_i) \cdot \mathcal{P}_{k}(\hat{P}_j) \nonumber \\
        & = \frac{1}{D^2} \sum_{i,j, k} e_{ij} \epsilon_{ik} \epsilon_{jk} \hat{P}_i \cdot \hat{P}_j.
    \end{align}
    To go ahead, let $\hat{P}_l = \hat{P}_i \hat{P}_j$, then we have 
    \begin{equation}
        \mathcal{P}_k(\hat{P}_l) = \mathcal{P}_k(\hat{P}_i) \mathcal{P}_k(\hat{P}_j),
    \end{equation}
    which means $\epsilon_{ik} \epsilon_{jk} = \epsilon_{lk}$.
    Note that for $l = 0$, all $\epsilon_{lk} = 1$, otherwise, $\epsilon_{lk} = \pm$ with equal probabilities, thus 
    \begin{equation}
        \sum_k \epsilon_{lk} = D^2 \delta_{l,0} = D^2 \delta_{i,j}.
    \end{equation}
    With this result, the twirling channel
    \begin{equation}
        \mathcal{E}_{\mathrm{twirl}} = \sum_{i} e_{ii} \mathcal{P}_i
    \end{equation}
    is Pauli diagonal.

    However, in practice, the error channel $\mathcal{E}$ company with the ideal gate $\mathcal{U}$, so our task is to twirl the error channel $\mathcal{E}$ in the noisy gate $\mathcal{U}_{\lambda} = \mathcal{E} \circ \mathcal{U}$ without changing the ideal gate $\mathcal{U}$.
    To do so, notice that the Clifford gate combined with T-gate forms the universal quantum gate set.
    For the Clifford gate $\mathcal{C}$, the twirling channel of its noisy realization is 
    \begin{equation} \label{eq: twirling}
        \mathcal{C}_{\lambda \mathrm{twirl}} = \frac{1}{|P|} \sum_{\mathcal{P} \in P} \mathcal{P} \circ \mathcal{C}_{\lambda} \circ \mathcal{P}',
    \end{equation}
    where $\mathcal{P}' = \mathcal{C}^{\dagger} \circ \mathcal{P} \circ \mathcal{C}$ are also Pauli gates.
    In this selection of $\mathcal{P}'$, the twirling channel of noisy gate is
    \begin{align}
        \mathcal{C}_{\lambda \mathrm{twirl}} & = \frac{1}{|P|} \sum_{\mathcal{P} \in P} \mathcal{P} \circ \mathcal{E} \circ \mathcal{C} \circ \mathcal{P}' \nonumber \\
        & = \frac{1}{|P|} \sum_{\mathcal{P} \in P} \mathcal{P} \circ \mathcal{E} \circ \mathcal{P} \circ \mathcal{C} \nonumber \\
        & = \mathcal{E}_{\mathrm{twirl}} \circ \mathcal{C}.
    \end{align} 
    For the T-gate $\mathcal{T}$, the construction is similar
    \begin{equation}
        \mathcal{T}_{\lambda \mathrm{twirl}} = \frac{1}{|P|} \sum_{\mathcal{P} \in P} \mathcal{P} \circ \mathcal{T}_{\lambda} \circ \mathcal{P}',
    \end{equation}
    while $\mathcal{P}' = \mathcal{T}^{\dagger} \circ \mathcal{P} \circ \mathcal{T}$ are Clifford gates instead.
    Note that if the error channel is so defined that on the ideal gate $\mathcal{U}_{\lambda} = \mathcal{U} \circ \mathcal{E}$, the twirling channel can be constructed similarly, with a small modification 
    \begin{equation}
        \mathcal{U}_{\lambda \mathrm{twirl}} = \frac{1}{|P|} \sum_{\mathcal{P} \in P} \mathcal{P}' \circ \mathcal{U}_{\lambda} \circ \mathcal{P},
    \end{equation} 
    where $\mathcal{P}' = \mathcal{U} \circ \mathcal{P} \circ \mathcal{U}^{\dagger}$.

\section{Effectiveness} \label{app: effectiveness}

    The failing probability is 
    \begin{align}
        P&\left(\overline{\chi_n^2}^{-1/2} \left\vert\chi_{ni} - \overline{\chi_n^2}^{1/2}\right\vert \geq \epsilon\right) \nonumber\\
        & = P\left(\chi_{ni} \overline{\chi_n^2}^{-1/2}- 1 \geq \epsilon\right) \nonumber \\
        & \ \quad + P\left(1 - \chi_{ni} \overline{\chi_n^2}^{-1/2} \geq \epsilon\right).
    \end{align}
    By exponential from of Chebyshev's inequality, and $\overline{\chi_n^2} = \bar{\chi}_n^2 + \Delta\bar{\chi}_n^2$, we have
    \begin{align}
        & P\left(\chi_{ni} \overline{\chi_n^2}^{-1/2}- 1 \geq \epsilon\right) \leq \inf_{\lambda>0} \mathbb{E} e^{\lambda \left(\chi_{ni} \overline{\chi_n^2}^{-1/2} -1 - \epsilon\right)}   \nonumber \\
        & \approx \inf_{\lambda>0} \exp\left[\frac{\Delta\bar{\chi}_n^2}{2 \overline{\chi_n^2}} \lambda^2 + \lambda \left(\frac{\bar{\chi}_n}{\overline{\chi_n^2}^{1/2}} - 1 - \epsilon\right) \right] \nonumber \\
        & \approx \inf_{\lambda>0} \exp\left[\frac{\Delta\bar{\chi}_n^2}{2 \bar{\chi}_n^2} \lambda^2 - \lambda \left(\epsilon + \frac{\Delta\bar{\chi}_n^2}{2 \bar{\chi}_n^2}\right) \right] \nonumber \\
        & = \exp\left(- \frac{\bar{\chi}_n^2}{2 \Delta\bar{\chi}_n^2}\epsilon^2 - \frac{1}{2} \epsilon\right),
    \end{align}
    where in the second line, we use the cumulant expansion to the second order, and in the third line, we keep to the first order of $\Delta\bar{\chi}_n^2$.
    Similarly, we have 
    \begin{align}
        & P\left(1 - \chi_{ni} \overline{\chi_n^2}^{-1/2} \geq \epsilon\right) \leq \inf_{\lambda>0} \mathbb{E} e^{\lambda \left(1 - \chi_{ni} \overline{\chi_n^2}^{-1/2} - \epsilon\right)}   \nonumber \\
        & \approx \inf_{\lambda>0} \exp\left[\frac{\Delta\bar{\chi}_n^2}{2 \overline{\chi_n^2}} \lambda^2 + \lambda \left(1 - \frac{\bar{\chi}_n}{\overline{\chi_n^2}^{1/2}} - \epsilon\right) \right] \nonumber \\
        & \approx \inf_{\lambda>0} \exp\left[\frac{\Delta\bar{\chi}_n^2}{2 \bar{\chi}_n^2} \lambda^2 - \lambda \left(\epsilon - \frac{\Delta\bar{\chi}_n^2}{2 \bar{\chi}_n^2}\right) \right] \nonumber \\
        & = \exp\left(- \frac{\bar{\chi}_n^2}{2 \Delta\bar{\chi}_n^2}\epsilon^2 + \frac{1}{2} \epsilon\right).
    \end{align}
    These give the result in Equation~(\ref{eq: effectiveness}).

    Then, we estimate the tolerant error $\epsilon$ in detail.
    By $\chi_{n_i} = \sum_{j} \epsilon_{ij} q_{nj}$, where $\epsilon_{ij}$ is the parity between Pauli operators $\hat{P}_i, \hat{P}_j$, the average is
    \begin{equation}
        \bar{\chi}_n = \mathbb{E}_i(\chi_{n_i}) = \sum_{j} q_{nj} \mathbb{E}_i(\epsilon_{ij}).
    \end{equation}
    For $j = 0$, $\epsilon_{ij} = 1$, while $j \neq 0$, $\epsilon_{ij} = \pm 1$ distributes with equal probability $\frac{1}{2}$, with the assumption that $\rho_i^2$ are, so the average
    \begin{equation}
        \mathbb{E}_i(\epsilon_{ij}) = \frac{1}{Z} \sum_i \rho_i^2 \epsilon_{ij} = \frac{1}{Z} \mathbb{E}_{\epsilon} (\epsilon \rho_{\epsilon}^2)
    \end{equation}
    where we treat $\rho_i^2$ as a random variable $\rho_{\epsilon}^2 = \sum_{\epsilon_i = \epsilon} \rho_i^2$ distributing over the parity $\epsilon$, and $Z = \sum_i \rho_i^2$.
    Without accurate knowledge of the state, we assume the distribution of $\rho_{\epsilon}^2$ is nearly symmetric, thus $\mathbb{E}_i(\epsilon_{ij}) \sim 0$.
    In this consideration, we have 
    \begin{equation}
        \bar{\chi}_n \sim q_{f0} = 1 - q_{\lambda}.
    \end{equation}
    For the second-order moment, 
    \begin{align}
        \overline{\chi_n^2} & = \mathbb{E}_i(\chi_{n_i}^2) = \sum_{j} q_{fj}^2 + \sum_{j\neq k} q_{nj} q_{nk} \mathbb{E}_i(\epsilon_{ij} \epsilon_{ik}) \nonumber \\
        & \sim \sum_{j} q_{nj}^2 = (1 - q_{\lambda})^2 + \sum_{j\neq 0} q_{nj}^2 \nonumber \\
        & \leq (1 - q_{\lambda})^2 + q_{\lambda}^2,
    \end{align}
    where the second line follows that when $j \neq k$, $\epsilon_{ij}$ and $\epsilon_{ik}$ are independent, and the third line is because $q_{nj} \leq q_{\lambda}$.
    Therefore, the parameter $\sigma$ is of the order
    \begin{equation}
        \sigma \sim \frac{q_{\lambda}}{1 - q_{\lambda}},
    \end{equation} 
    and the tolerant error
    \begin{equation}
        \epsilon \sim 2 q_{\lambda} \sqrt{\frac{2 - \delta}{4 - 8 q_{\lambda} + 3 q_{\lambda}^2}} \sim \sqrt{2} \lambda + O(\lambda^2).
    \end{equation}

    We calculate the tolerant error of relative bias for extrapolation to pure state, $s_{n_0} = 0$. 
    With $\overline{\chi_{n_0}^2} = \mathbb{E}(\chi_{n_0 i}^2) = 1$, and by Equation~(\ref{eq: second}), we have 
    \begin{align}
        \overline{\chi_{n_0}^2} & = \mathbb{E}\left[\chi_{fi}^{2 (2 n_0 - 1)} e^{2 (n_0 - 1) \lambda^2 \omega_i}\right] \nonumber \\
        & \approx \exp\left[2 (n_0 - 1) \lambda^2 (\bar{\omega} + (n_0 - 1) \lambda^2 \Delta \bar{\omega}^2)\right],
    \end{align}
    where we expect $n_0 \sim \frac{1}{2}$ and use cumulant expansion to second order.
    Thus, the effective noise-free point is 
    \begin{equation}
        n_0 \approx 1 - \frac{\bar{\omega}}{\lambda^2 \Delta \bar{\omega}^2}.
    \end{equation}
    It follows that $\bar{\chi}_{n_0} = e^{- \frac{\lambda^2 \bar{\omega}}{2}(1 - n_0)}$, and the relative bias is
    \begin{equation}
        \epsilon \sim \sqrt{2} \sigma  \sim \sqrt{2} (1 -n_0) \lambda^2 \Delta \bar{\omega} + O(\lambda^4),
    \end{equation}

    For routine ZNE method, the relative bias is $\left|e^{- \lambda^2 \omega_i/2} - 1\right|$, and the failing probability is calculated as
    \begin{equation}
        P\left(\left|\cdot\right| \geq \epsilon\right) \leq 2 \exp\left(- \frac{2 \epsilon^2}{\lambda^4 \Delta \bar{\omega}^2} - \frac{\bar{\omega}^2}{2 \Delta \bar{\omega}^2}\right) \cosh \frac{2 \bar{\omega} \epsilon}{\lambda^2 \Delta \bar{\omega}^2}.
    \end{equation}
    The tolerant error is bounded as 
    \begin{equation}
        \epsilon \leq \frac{\sqrt{2} \lambda^2 \Delta \bar{\omega}}{2} \sqrt{\frac{1 - {\bar{\omega}^2}/{\Delta \bar{\omega}^2} - \delta/2}{1 - 2 (1-n_0)^2 \lambda^4 \Delta \bar{\omega}^2}} \sim \frac{\sqrt{2}}{2} \lambda^2 \Delta \bar{\omega},
    \end{equation}

    For the VD/ESD method with $2$-replica, with $\bar{\chi}_n \sim 1 - q_{\lambda} \sim 1- \lambda$ and $\Delta \bar{\chi}_n \sim q_\lambda \sim \lambda$, the failing probability is calculated in the similar way,
    \begin{align}
        P \left(\left|\cdot\right| \geq \epsilon \right) & \lesssim 2 \exp\left[- \frac{\epsilon^2}{2 \lambda^2} - \frac{D - 2}{2D} \right] \nonumber \\
        & \ \quad \times \cosh\left[\frac{D - 1}{D}\left(\frac{1}{\lambda} - \frac{2D + 1}{D}\right) \epsilon \right] \\
        & \sim 2 e^{- \frac{\epsilon^2}{2 \lambda^2} - \frac{1}{2}} \cosh \frac{\epsilon}{\lambda} \nonumber \\
        & \sim 2 e^{- \frac{1}{2}} \left(1 - \frac{\epsilon^4}{12 \lambda^4}\right) + O(\epsilon^6)
    \end{align} 
    where $\left|\cdot\right|$ stands for $\left|\frac{D \chi_{ni}}{(D - 1) \overline{\chi_n^2} + 1} - 1\right|$, and we assume $\lambda \ll 1$ and $D \gg 1, \epsilon \sim \lambda$ in the third line.
    This shows that the tolerant error 
    \begin{equation}
        \epsilon \sim \sqrt[4]{12} \lambda.
    \end{equation}

\section{Efficiency} \label{app: efficiency}
To calculate the overhead $C_{em}$, we consider the variance of the estimator.
Since the estimator $O_{em}$ is not linear, the variance is not defined rigidly.
But up to linear order, we can derive it approximately.
\begin{align}
    & \mathrm{Var} [O_{em}] \sim  \ \mathrm{Var}[\Delta F] \\
    & = \mathrm{Var}\left[\sum_i \left(\frac{\partial F}{\partial O_i} \Delta O_i + \frac{\partial F}{\partial p_i} \Delta p_i\right)\right] \nonumber \\
    & \leq \left(\sum_i \left|\frac{\partial F}{\partial O_i}\right| \mathrm{Var}^{1/2}[O_i] + \left|\frac{\partial F}{\partial p_i}\right| \mathrm{Var}^{1/2}[p_i]\right)^2 \nonumber\\
    & \lesssim \left(\mathrm{Var}[O] \sum_i \left|\frac{\partial F}{\partial O_i}\right|  + 2 \mathrm{Var}[p] \sum_i \left|\frac{\partial F}{\partial p_i}\right|\right)^2.
\end{align}
According to the law of large numbers (LLN), the overhead is given by
\begin{align}
    C_{em} & \equiv \frac{N^{\epsilon}(O_{em})}{N^{\epsilon}(O)} = \frac{\mathrm{Var} [O_{em}]}{\mathrm{Var}[O]} \nonumber\\
    & \lesssim \left(\sum_i \left|\frac{\partial F}{\partial O_i}\right| + \frac{\mathrm{Var}[p]}{\mathrm{Var}[O]} \sum_i \left|\frac{\partial F}{\partial p_i}\right|\right)^2 \nonumber \\
    & \sim \left(\sqrt{C_{em}^{\textrm{ZNE}}} + \frac{\mathrm{Var}[p]}{\mathrm{Var}[O]} \sum_i \left|\frac{\partial F}{\partial p_i}\right|\right)^2,
\end{align}
where $N^{\epsilon}(O_{em})$ and $N^{\epsilon}(O)$ is the number of shots to measure the mitigated estimator $O_{em}$ and the raw estimator $O$ up to $\epsilon$ precision correspondingly.
Here, we assume that the overhead of routine ZNE method $C_{em}^{\textrm{ZNE}} \sim \sum_i \left({\partial F}/{\partial O_i}\right)^2$ approximates to the observable expectation part of the overhead of pZNE.
The reason is that the purity influences the mitigated estimator $O_{em}$ via the error-rate-like index $s$, thus the fitting model $O(s)$ is similar to $O(x)$ of ZNE.

In the second term of the overhead $C_{em}$, the part depending on $F$ is correlated to the specific fitting model.
We only consider the variance $\mathrm{Var}[p]$ of purity in the latter.
The variance $\mathrm{Var}[p]$ depends on the method to measure the purity, and as we previously discussed, there are two kinds of methods to measure the purity.
One is the replica measurement method, which is based on the Equation~(\ref{eq: S}).
Another one is the QST-like method, such as QST and classical shadow, which measure the expectation of all Pauli operators $\braket{\hat{\sigma}_{\alpha}}$, where $\alpha \in \{1, 2, \dots, 4^{L}-1\}$, and $L$ is the system size.
The definition of the purity is
\begin{equation}
    p = \frac{1}{2^L} \left(1 + \sum_{\alpha} \braket{\hat{\sigma}_{\alpha}}^2\right).
\end{equation}
Then, we calculate the overhead of these two different measurement methods respectively.

For the replica measurement method, the variance of purity is
\begin{equation}
    \mathrm{Var}[p] = \frac{1}{N} \mathrm{Var}\left[\hat{S}^{(2)}\right] = \frac{1 - p^2}{N}.
\end{equation}
Next, if we assume the observable of interest is a Pauli operator, we have $\mathrm{Var}[O] = \frac{1 - O^2}{N}$, thus
\begin{equation}
    \frac{\mathrm{Var}[p]}{\mathrm{Var}[O]} = \frac{1 - p^2}{1 - O^2}.
\end{equation}
When the error rate $x \rightarrow 1$, we have $p \rightarrow 1$, if $O \not \rightarrow 1$, then the second term of the overhead $C_{em}$ has less influence, and we have
\begin{equation}
    C_{em} \sim C_{em}^{\textrm{ZNE}}.
\end{equation}
In the contrary, if $O \rightarrow 1$ also, we have
\begin{equation}
    \frac{\mathrm{Var}[p]}{\mathrm{Var}[O]} \rightarrow \frac{1 - p}{1 - O} \rightarrow \frac{2 K_A}{K_B} \equiv 2 \kappa < \infty,
\end{equation}
where $K_A \equiv \sum_a A_a(0) \Re(k_a)$, and $K_B \equiv \sum_a B_a(0) k_a = \sum_a \tilde{B}_a(0) \Re(k_a)$.
The second equality of $K_B$ is because that $O = \braket{\hat{O}}$ is real.
Then the overhead is
\begin{equation}
    C_{em} \lesssim \left(\sqrt{C_{em}^{\textrm{ZNE}}} + 2 \kappa \sum_i \left|\frac{\partial F}{\partial p_i}\right|\right)^2.
\end{equation}

For the QST-like measurement method, we assume that the observable $\hat{O}$ is a Pauli operator located on $l_0$ qubits, namely $\hat{O}$ is not identity $\hat{I}$ exactly on $l_0$ qubits, and we call it the $l_0$-weight Pauli operator in the following.
Then, the variance of purity of system with $L$ qubits is
\begin{align}
    \mathrm{Var}[p] & = \mathrm{Var}\left[\sum_{\alpha}\frac{\partial p}{\partial \sigma_{\alpha}} \Delta \sigma_{\alpha}\right] = \sum_{\alpha} \left(\frac{\partial p}{\partial \sigma_{\alpha}}\right)^2 \mathrm{Var}[\sigma_{\alpha}] \nonumber \\
    & = \frac{1}{4^{L-1}} \sum_{\alpha} \braket{\hat{\sigma}_{\alpha}}^2 \mathrm{Var}[\sigma_{\alpha}].
\end{align}
We measure the $3^L$ different $L$-weight Pauli operators with the same number $N_L$ of shots to perform the QST-like method.
The number of shots used to estimate the $l$-weight Pauli operator shorter than $L$ is $N_l = 3^{L-l} N_L$.
Thus, by LLN, the variance of the $l$-weight Pauli operator is
\begin{equation} \label{eq: var}
    \mathrm{Var}[\sigma_{|\alpha| = l}] \sim \frac{N_{l_0}}{N_l} \mathrm{Var}[\sigma_{|\alpha| = l_0}] = 3^{l - l_0} \mathrm{Var}[O].
\end{equation}
Dividing the summation of $\alpha$ into the summation of weight $|\alpha| = l$, we have
\begin{align}
    \mathrm{Var}[p] & = \frac{1}{4^{L-1}} \sum_{l>0} 3^l C_L^l \overline{\braket{\hat{\sigma}_{|\alpha| = l}}^2} \mathrm{Var}[\sigma_{|\alpha| = l}] \nonumber \\
    & \sim \frac{\mathrm{Var}[O]}{4^{L-1}} \sum_{l>0} 3^{2l - l_0}C_L^l \overline{\braket{\hat{\sigma}_{|\alpha| = l}}^2} \nonumber \\
    & \lesssim \frac{\mathrm{Var}[O]}{4^{L-1} 3^{l_0}} \sum_{l>0} 3^{2l}C_L^l = \frac{10^L -1}{4^{L-1} 3^{l_0}} \mathrm{Var}[O],
\end{align}
where $\overline{\braket{\hat{\sigma}_{|\alpha| = l}}^2} \equiv \frac{1}{3^l C_L^l} \sum_{|\alpha|=l} \braket{\hat{\sigma}_{\alpha}}^2 \leq 1$.
Therefore, the overhead is bounded as
\begin{equation}
    C_{em} \lesssim \left(\sqrt{\tilde{C}_{em}^{\textrm{ZNE}}} + \frac{10^L -1}{4^{L-1} 3^{l_0}} \sum_i \left|\frac{\partial F}{\partial p_i}\right|\right)^2.
\end{equation}

\section{The Detail of Numerical Simulation} \label{app: simulation}
    
    \subsection{Data Process}

    For the extrapolation, we substitute the gates with the folded gates with odd numbers $1,3$ and $5$.
    In addition, we also simulate the case without any gate to separate the error of the circuit and the error of preparation and measurement. 
    The circuit with layers from $1$ to $20$ is simulated $10$ times.
    After each circuit with given layers, we perform the projective measurement on each qubit based on all three Pauli operators.
    Each Pauli operator is measured with $2,000$ shots.
    
    With the measurement outcome, we estimate the expectations of all Pauli operators of the two-qubit system, which is used to calculate the purity of states.
    For the operator $\hat{Z}_0$, it is estimated from $6,000$ shots, while the purity of the system is estimated from $18,000$ shots.
    For a given number of layers, there are results of four different circuits, whose layer is $1,3$ and $5$ folded gate.
    The estimator of ZNE is extrapolated to the noise-free gate from the expectation $\braket{\hat{Z}_0}$ of $1,3$ and $5$ folded gate, with the exponential fitting model
    \begin{equation}
        \braket{\hat{Z}_0}_n = A e^{-k n} + B ,
    \end{equation} 
    where $(A, k, B)$ are the parameters to be determined, and $x$ is the number of folding, namely $1,3$ and $5$.
    The estimator of pZNE is extrapolated from the expectation $\braket{\hat{Z}_0}(x)$ of $1,3$ and $5$ folded gates, and the purity $p(x)$ of $1,3$ and $5$ folded gates.
    The fitting model is 
    \begin{equation}
        p = A \braket{\hat{Z}_0}^k + C,
    \end{equation}
    and the estimator is
    \begin{equation}
        \braket{\hat{Z}_0}_{\mathrm{pZNE}} = \left(\frac{1 - C}{A}\right)^{1/k}.
    \end{equation}
    We fit the parameters $A$, $k$, and $B$ by the least square method.

    For the VD/ESD method, we use the state estimator with $M = 2$ replica
    \begin{equation}
        \rho_{\mathrm{p}} = \frac{\rho_{n}^2}{\mathrm{Tr} \rho_{n}^2},
    \end{equation}
    where $\rho_{n}$ is the noisy state of the qubit $q_0$.
    The estimator of expectation is 
    \begin{equation}
        \braket{\hat{Z}_0}_{\mathrm{p}} = \mathrm{Tr}(\hat{Z}_0 \rho_{\mathrm{p}}) = \frac{\mathrm{Tr}[\hat{Z}_0 \rho_{n}^2]}{\mathrm{Tr} \rho_{n}^2}.
    \end{equation}
    It can be verified that $\mathrm{Tr}[\hat{Z}_0 \rho_{n}^2] = \mathrm{Tr}[\hat{Z}_0 \rho_{n}] = \braket{\hat{Z}_0}_{n}$, thus
    \begin{equation}
        \braket{\hat{Z}_0}_{\mathrm{p}} = \frac{\braket{\hat{Z}_0}_{n}}{p_n}.
    \end{equation}
    The estimator of the modified purification method is 
    \begin{equation}
        \braket{Z_0}_{\mathrm{m}} = \braket{\hat{Z}_0}_{n} \sqrt{\frac{D - 1}{D p_n - 1}}.
    \end{equation}

    \subsection{Pauli Errors Sampling}

    We randomly sample the Pauli diagonal errors $\mathcal{E} =\sum_i q_i \mathcal{P}_i$, where $\sum_i q_i = 1$, with a fixed error probability $p_{\lambda} = 1 - p_0$.
    Therefore, all these errors form located on a $4^n - 2$ dimension projective space $P\mathbb{R}^{4^n - 2}$.
    By stereographic projection, the projective space is diffeomorphic to Riemann sphere $\mathbb{S}^{4^n - 2}$.
    The errors are sampled uniformly on this sphere.

    First, we uniformly sample $4^n - 1$ number $r_i$ from $[0, 1]$.
    The points formed from them are discarded if it is located out of the inscribed sphere, otherwise, the point is normalized to the surface of the sphere.
    Then, the point on spherical surface $\mathbb{S}^{4^n - 2}$ is map to the projective space $P\mathbb{R}^{4^n - 2}$ by stereographic projection.
    The point on projective space which has negative components is also discarded.
    The left point is normalized to the given error probability $p_{\lambda} = 1 - p_0$.

    The parameters of error model used to obtain the results of Figure~\ref{fig: run_pauli}(a),~(b) and~(c) is shown in Table~\ref{tab: pauli_error}.
    \begin{table}
        \centering
        \begin{tabular}[htbp]{c|cccc}
            \hline
            $p_i$ & $I_1$ & $X_1$ & $Y_1$ & $Z_1$ \\ \hline
            $I_0$ & $9.50 \times 10^{-1}$ & $6.24\times 10^{-3}$ & $5.87\times 10^{-3}$ & $3.61\times 10^{-3}$\\
            $X_0$ & $3.22\times 10^{-3}$ & $1.64\times 10^{-3}$ & $5.19\times 10^{-3}$ & $3.80\times 10^{-4}$ \\
            $Y_0$ & $4.15\times 10^{-3}$ & $6.89\times 10^{-3}$ & $4.01\times 10^{-3}$ & $4.00\times 10^{-4}$ \\
            $Z_0$ & $7.50\times 10^{-4}$ & $2.04\times 10^{-3}$ & $3.47\times 10^{-3}$ & $2.14\times 10^{-3}$ \\
            \hline
            \end{tabular}
            \caption{The probability $q_i$ of Pauli basis $\mathcal{P}_i$ of Pauli diagonal error model in simulations of Figure~\ref{fig: run_pauli}(a),~(b), and~(c). The leftmost column denotes the Pauli operations on qubit $q_0$, and the top row denotes the Pauli operations on $q_1$. } \label{tab: pauli_error}
    \end{table}

    \subsection{Pauli Twirling of CNOT Gate}
    
    For our circuits in the experiment, Pauli twirling is performed for the whole circuit.
    The Pauli gates $\mathcal{P}_1, \mathcal{P}_2$ in Figure~\ref{fig: exper}(a) run over the $16$ elements of Pauli group, and the Pauli gates $\mathcal{P}_3, \mathcal{P}_4$ are selected by the Equation~\ref{eq: twirling} accordingly. 

    If $L$ is even, then the ideal circuit is identity, and $\mathcal{P}_3 = \mathcal{P}_1, \mathcal{P}_4 = \mathcal{P}_2$.
    Otherwise, when $L$ is odd, $\mathcal{P}_3, \mathcal{P}_4$ can be calculated by Equation~\ref{eq: propogation}.
    All the $16$ instances are shown in Table~\ref{tab: cnot_twirling}.

    \begin{table}
        \centering
        \begin{tabular}[htbp]{c|cccc}
            \hline
            {$\mathcal{P}_3 \mathcal{P}_4$} & $I_2$ & $X_2$ & $Y_2$ & $Z_2$ \\ \hline
            $I_1$ & $I_1 I_2$ & $I_1 X_2$ & $Z_1 Y_2$ & $Z_1 Z_2$\\
            $X_1$ & $X_1 X_2$ & $X_1 I_2$ & $Y_1 Z_2$ & $Y_1 Y_2$ \\
            $Y_1$ & $Y_1 X_2$ & $Y_1 I_2$ & $X_1 Z_2$ & $X_1 Y_2$ \\
            $Z_1$ & $Z_1 I_2$ & $Z_1 X_2$ & $I_1 Y_2$ & $I_1 Z_2$ \\
            \hline
            \end{tabular}
            \caption{Pauli twirling of CNOT gate. The leftmost column denotes $\mathcal{P}_1$, and the top row denotes $\mathcal{P}_2$. } \label{tab: cnot_twirling}
    \end{table}

\section{Device Information on Quafu Cloud Platform} \label{app: experiment}
\begin{figure}[t]
        \centering
        \includegraphics[width=0.8\textwidth]{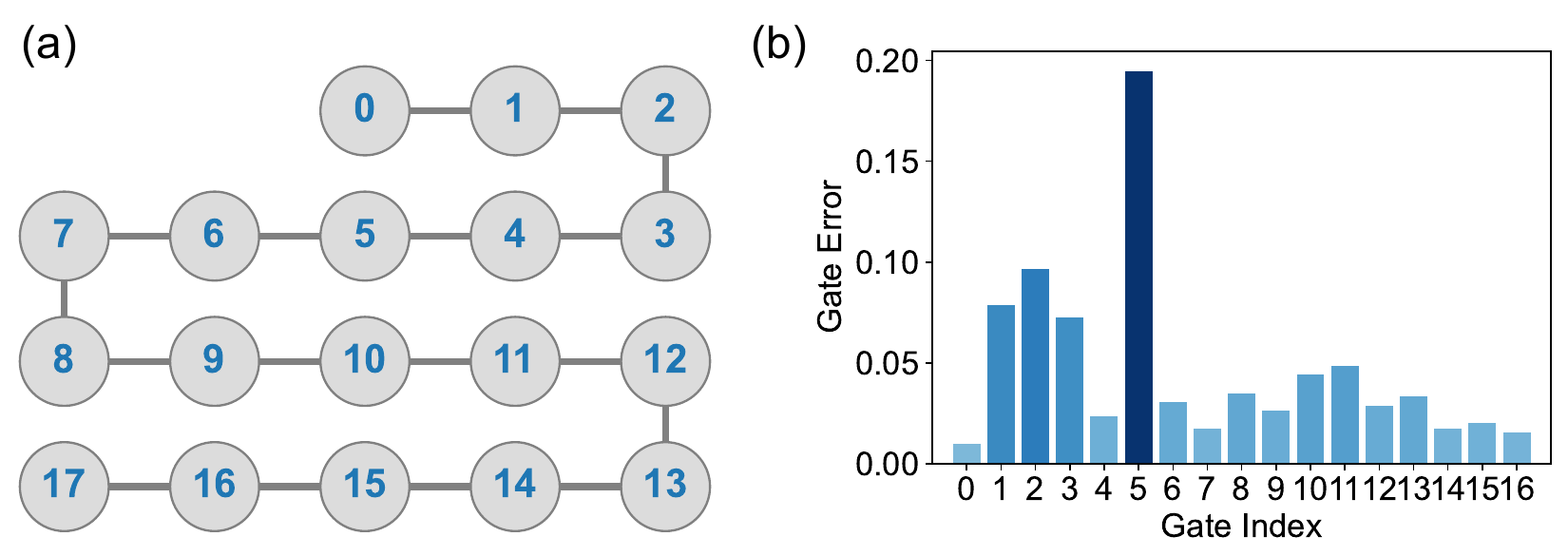}
        \caption{The layout of the chip and the error rates of CZ gates.}
        \label{fig: chip_info}
    \end{figure}

    In Sec.~\ref{sec: simu}, we use the ScQ-P18 device on Quafu cloud platform for experimental verification. Here we utilize the open-source Python SDK, namely PyQuafu, to write the experimental code for running the circuits. The layout of the ScQ-P18 device and the error rates of CZ gates are shown in Figure~\ref{fig: chip_info}. We use qubits 4 and 5 on ScQ-P18 for the two-qubit demonstration. The XEB fidelity of single-qubit gates is about 0.9947 for qubit 4, 0.9972 for qubit 5, and their CZ gate is about 0.966. More information about the two used qubits can be found in Table~\ref{tab: paras}.    

\begin{table}
\centering
\begin{tabular}[htbp]{ccc}
    \hline
    Qubit index & 4 & 5 \\[.06cm] \hline
        \\[-.4cm]
        Qubit frequency, $f_{10}$ ($\mathrm{GHz}$) & 4.930 & 4.465 \\[.06cm]
        Readout frequency, $f_{r}$ ($\mathrm{GHz}$) & 6.713 & 6.693 \\[.06cm]
        Anharmonicity, $\eta$ ($\mathrm{MHz}$) & -204.8 & -191.7 \\[.06cm]
        Relaxation time, $T_{1}$ ($\mu\mathrm{s}$) & 23.37 & 31.03 \\[.06cm]
        Coherence time, $T_{2}$ ($\mu\mathrm{s}$) & 2.47 & 1.51 \\[.06cm]
        Readout fidelity of 
state $\ket{0}$, $F_{0}$ & 0.9524 & 0.9109 \\[.06cm]
        Readout fidelity of 
state $\ket{1}$, $F_{1}$ & 0.9025 & 0.8647 \\[.06cm]
    \hline
\end{tabular}
    \caption{The parameters of the two used qubits.} \label{tab: paras}
\end{table}

    Here we mitigate the readout error of the experimental results on the cloud platform. In specific, we first assume the ideal and noisy measurements are in bases $\ket{x}$ and $\ket{E_x}$, respectively. According to the Bayesian readout correction, the relation between the noisy measurement operator $M_{x}:=\ket{E_x}\bra{E_x}$ and the ideal operator $P_x:= \ket{x}\bra{x}$ can be expressed as
    \begin{equation}
        M_{x}=\sum_{y}R_{xy}P_{y},
    \end{equation}
    where the matrix $R$ can be evaluated from the probability of experimental outcome when performing the noisy measurement on the prepared states $\ket{x}$, namely
    \begin{equation}\label{eq:fid_matrix}
        R_{xy}=\mathrm{Tr}(M_{x} P_{y}).
    \end{equation}
    In the single-qubit case, the matrix $R$ is constructed by the readout fidelities:
    \begin{equation}
        R=\left(\begin{array}{cc}
        F_0 & 1-F_1 \\
        1-F_1 & F_1
    \end{array} \right)
    \end{equation}
    where $F_0$ and $F_1$ are the readout fidelities of states $\ket{0}$ and $\ket{1}$. Ignoring the readout crosstalk, the multi-qubit $R$ matrix can be reduced to the direct product of each single-qubit $R$ matrix.
    
    With the knowledge of matrix $R$, we can construct the ideal measurement outcome by using its inverse
    \begin{equation}
        P_{x} = \sum_{y} (R^{-1})_{xy} M_{y},
    \end{equation}
    and the probability $p_x = \mathrm{Tr}(P_x \rho)$ of some state $\rho$ in state $\ket{x}$ is thus
    \begin{equation}
        p_x = \sum_{y} (R^{-1})_{xy} q_y,
    \end{equation}
    where $q_x = \mathrm{Tr}(M_{x} \rho)$ is the probability of this state in $\ket{E_x}$. In addition, we also use the least square method to limit the mitigated probabilities, so that all of them meet the non-negative and normalization conditions.

\medskip
\textbf{Acknowledgements} \par 
This work was supported by National Natural Science Foundation of China (Grants Nos.92265207, T2121001, 11934018, 12122504), the Innovation Program for Quantum Science and Technology (Grant No. 2021ZD0301800), and Beijing Natural Science Foundation (Grant No. Z200009).
We also acknowledge the support from the Synergetic Extreme Condition User Facility (SECUF) in Beijing, China.

\medskip
\textbf{Conflict of Interest} \par 
The authors declare no conflict of interest.

\medskip
\textbf{Data Availability Statement} \par 
The data that support the findings of this study are available from the corresponding author upon reasonable request. 

\medskip

%



\begin{figure}
\textbf{Table of Contents}\\
\medskip
  \includegraphics{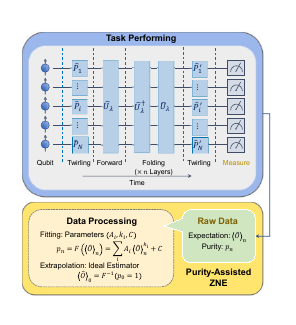}
  \medskip
  \caption*{Zero noise extrapolation (ZNE) is an error mitigation method that amplifies and extrapolates noise to a noise-free point, yet it relies on assumptions about the error model. This paper introduces a modified method, termed purity-assisted ZNE (pZNE), which addresses these limitations by utilizing the Pauli twirling technique and leveraging the purity of the noisy output state.}
\end{figure}

\end{document}